\documentclass[floats,floatfix,showpacs,amssymb,prd,twocolumn,superscriptaddress,nofootinbib]{revtex4-1}

\usepackage{amssymb,amsmath,verbatim,mathtools,needspace,enumitem,etoolbox,graphicx,microtype}
\usepackage{multirow}
\usepackage[dvipsnames, usenames]{xcolor}

\definecolor{linkcolor}{rgb}{0.0,0.3,0.5}
\usepackage[unicode, colorlinks=true, linkcolor=linkcolor, citecolor=linkcolor, filecolor=linkcolor,urlcolor=linkcolor, pdfusetitle]{hyperref}
\usepackage[all]{hypcap}
\usepackage[T1]{fontenc}
\usepackage[utf8]{inputenc}

\usepackage[utf8]{inputenc}
\usepackage{graphicx,amssymb,amsmath,amsthm,amsfonts,epsfig,times,natbib}
\usepackage{placeins}

\usepackage{epstopdf}
\usepackage{tensor}
\usepackage{amsbsy}
\usepackage{bm,url}
\usepackage{float}
\usepackage{dcolumn}
\usepackage{latexsym}
\usepackage{rotating}
\usepackage{longtable}
\usepackage{enumerate}
\usepackage{booktabs}
\usepackage[utf8]{inputenc}
\newcommand{\msun}{\ensuremath{M_\odot}}
\newcommand{\beq}{\begin{eqnarray}}
\newcommand{\eeq}{\end{eqnarray}} 
\newcommand{\ba}{\begin{align}}
\newcommand{\ea}{\end{align}}
\newcommand{\be}{\begin{equation}}
\newcommand{\ee}{\end{equation}}
\begin{document}

\title{Follow-up signals from superradiant instabilities of black hole merger remnants}

\author{Shrobana Ghosh}
\email{sghosh2@go.olemiss.edu}
\affiliation{Department of Physics and Astronomy, The University of  Mississippi, University, MS 38677, USA}
\author{Emanuele Berti}
\email{berti@jhu.edu}
\affiliation{Department of Physics and Astronomy, Johns Hopkins University, 3400 N. Charles Street, Baltimore, MD 21218, USA}
\affiliation{Department of Physics and Astronomy, The University of  Mississippi, University, MS 38677, USA}
\author{Richard Brito}
\email{richard.brito@aei.mpg.de}
\affiliation{Dipartimento di Fisica, ``Sapienza'' Universit\`a di Roma \& Sezione INFN Roma1, Piazzale Aldo Moro 5, 00185, Roma, Italy}
\affiliation{Max Planck Institute for Gravitational Physics (Albert Einstein Institute), Am M\"{u}hlenberg 1, Potsdam-Golm, 14476, Germany}
\author{Mauricio Richartz}
\email{mauricio.richartz@ufabc.edu.br}
\affiliation{Centro de Matem\'atica, Computa\c{c}\~ao e Cogni\c{c}\~ao,
Universidade Federal do ABC (UFABC), 09210-170 Santo Andr\'e, S\~ao Paulo, Brazil}
%

\date{\today}

\begin{abstract}
Superradiant instabilities can trigger the formation of bosonic clouds around rotating black holes. If the bosonic field growth is sufficiently fast, these clouds could form shortly after a binary black hole merger. Such clouds are continuous sources of gravitational waves whose detection (or lack thereof) can probe the existence of ultralight bosons (such as axion-like particles) and their properties. Motivated by the binary black hole mergers seen by Advanced LIGO so far, we investigate in detail the parameter space that can be probed with continuous gravitational wave signals from ultralight scalar field clouds around black hole merger remnants with particular focus on future ground-based detectors (A+, Voyager and Cosmic Explorer). We also study the impact that the confusion noise from a putative stochastic gravitational-wave background from unresolved sources would have on such searches and we estimate, under different astrophysical priors, the number of binary black-hole merger events that could lead to an observable post-merger signal. Under our most optimistic assumptions, Cosmic Explorer could detect dozens of post-merger signals.
\end{abstract}

\maketitle

\section{Introduction}
The detection of gravitational waves (GWs) from several binary black-hole (BH) mergers has opened up new opportunities to search for ultralight bosons predicted by extensions of the Standard Model. One such candidate is the axion, an elementary particle that was first proposed about 40 years ago to solve the strong CP problem of QCD~\cite{Wilczek:1977pj}. In the ``string axiverse'' scenario, pseudoscalar fields with axion-like properties generically arise in string theory compactifications as Kaluza–Klein zero modes of antisymmetric tensor fields, with potentially observable astrophysical consequences~\cite{Arvanitaki:2009fg}. Axion-like particles have also been considered as cold dark matter candidates~\cite{Bertone:2004pz}.
Decades of unsuccessful laboratory, astrophysical and cosmological searches for axions have put stringent bounds on their masses and interaction potentials~\cite{Marsh:2015xka,Poulin:2018dzj}.

GW astronomy can either detect axion-like fields, or set stringent bounds on their masses~\cite{Arvanitaki:2010sy,Arvanitaki:2014wva,Brito:2014wla,Arvanitaki:2016qwi,Baryakhtar:2017ngi,Brito:2017wnc,Brito:2017zvb,Baumann:2018vus,Hannuksela:2018izj,Isi:2018pzk}. One of the main ideas behind ultralight boson searches with GW detectors relies on the superradiant instability of rotating (Kerr) BHs~\cite{Arvanitaki:2010sy,Brito:2015oca}.

Consider a Kerr BH of mass $M$ and dimensionless spin $\chi=a/M$ and a boson of mass $m_s$, or reduced mass $\mu=m_s/\hbar$ (we use geometrical units, $G=c=1$). At the linear level, the radial potential describing perturbations induced by massive fields in asymptotically flat BH backgrounds has a potential well, and therefore time-decaying quasibound states. If the BH is rotating, however, {\em superradiant} modes of bosonic fields may be confined within the potential well, generating unstable quasibound states. The superradiant instability of these states is strongest when $\chi \sim 1$ and when the Compton wavelength of the boson is comparable to the BH Schwarzschild radius, i.e. $2M \mu \sim 1$~\cite{Detweiler:1980uk,Dolan:2007mj}. 
It has long been known that a rotating BH can lose at most $\sim$ 29\% of its mass~\cite{Christodoulou:1972kt}; in fact, more recent numerical work shows that superradiant instabilities can extract at most $\sim$ 10\% of a BH's mass~\cite{Herdeiro:2017phl,East:2017ovw,East:2018glu}.
When the instability operates, small (classical or quantum)
fluctuations of the bosonic field are allowed to grow in time. The
final outcome is a boson cloud around a more slowly spinning BH that
acts as a continuous GW source, with possibly important astrophysical
consequences.\footnote{This assumes that the boson field is described
  by a noninteracting massive scalar. If self-interactions and/or
  couplings to matter are strong enough, the overall dynamics will be
  more
  complicated~\cite{Yoshino:2012kn,Rosa:2017ury,Boskovic:2018lkj,Ikeda:2018nhb} and
  gravitational radiation might be suppressed. In particular,
  self-interactions are expected to be important for string-motivated
  axions~\cite{Arvanitaki:2009fg}, but we do not consider this case here.}

Astronomical evidence and theoretical predictions suggest that spinning BHs should be common in the Universe. Thermal continuum fitting and inner disk reflection models yield observational evidence for the existence of stellar-mass BHs with dimensionless spins as high as $\chi \sim 0.98$. In fact, BH spin measurements of X-ray binaries suggest that the mass interval $\sim [6\times 10^{-13},10^{-11}]$~eV is ruled out for non-interacting massive scalar fields~\cite{Arvanitaki:2014wva,Cardoso:2018tly}. However, these measurements are affected by systematic uncertainties (see e.g. Fig.~11 of~\cite{Krawczynski:2018fnw} for a comparison of the two methods for six stellar-mass BH systems). Besides, BH spin estimates using electromagnetic emission depend on the history of the binary. Any studies of superradiant instabilities using electromagnetic estimates of the spin are inevitably model-dependent, because both accretion and superradiance change the BH mass and spin.

Fortunately, nature gave us cleaner systems to study superradiant instabilities of rotating BHs. The ten binary BH merger candidates observed so far by the LIGO/Virgo collaboration provide relatively precise and unbiased measurements of the remnant BH mass $M$ and dimensionless spin $\chi$~\cite{LIGOScientific:2018mvr}, which rely only on general relativity being correct and are unaffected by accretion modeling systematics. The BH merger remnants observed in the first two LIGO/Virgo observing runs O1 and O2 have spins $0.66\lesssim \chi \lesssim 0.81$ (cf. Table~\ref{tab:ligofactsheets}), and we know with great accuracy when these rotating BHs were formed.

This begs the question: if ultralight bosons with $2M\mu  \sim 1$ exist in our Universe, so that the superradiant instability is effective, is the growth time of the cloud short enough (and is the superradiant GW signal strong enough) that GW detectors could carry out {\em follow-up observations} of the continuous GWs emitted by the boson cloud/BH system post-merger? This question was first raised in Ref.~\cite{Arvanitaki:2016qwi} and recently studied in more detail in Ref~\cite{Isi:2018pzk}, where it was shown that future GW detectors such as Cosmic Explorer could detect such sources out to $\sim 10$ Gpc, while Advanced LIGO at design sensitivity will reach distances of $\sim 100$ Mpc. In this paper we complement and extend those works by studying in more detail the parameter space that we will be able to probe with continuous post-merger GW signals emitted by the BH/cloud system. We consider the expected Advanced LIGO/Virgo design sensitivity~\cite{LIGOcurve} as well as future ground-based detectors that are expected to be operational in the next few years, including planned technological improvements within the current LIGO facilities (A+~\cite{Apluscurve} and Voyager~\cite{Voyagercurve}) as well as Cosmic Explorer, a 40~km design requiring new facilities~\cite{2017CQGra..34d4001A}.

The plan of the paper is as follows. In Sec.~\ref{sec:setup} we briefly review the BH/boson cloud model and the method used to compute the GW signal emitted by these sources.   
In Sec.~\ref{sec:constraints} we study the parameter space that we will be able to probe with post-merger GW signals. Following~\cite{Brito:2017wnc}, we also study how a stochastic background from all the unresolved BH/cloud sources would affect the detection of continuous post-merger GW signals. In Sec.~\ref{sec:rates} we extend previous studies~\cite{Arvanitaki:2016qwi} by computing the number of expected binary BH mergers that could lead to a detectable post-merger GW signal for Cosmic Explorer. Finally, in Sec.~\ref{sec:conclusions} we summarize our findings and discuss possible future improvements.

\begin{table}[t!]
\begin{tabular}{ccc}
\hline
\hline
Event & $M $ & $\chi$\\
\hline
\hline
\addlinespace[1mm]
GW150914~& $63.1^{+3.3}_{-3.0}$ & $0.69^{+0.05}_{-0.04}$\\
\addlinespace[1mm]
GW151012~ & $35.7^{+9.9}_{-3.8}$ & $0.67^{+0.13}_{-0.11}$\\
\addlinespace[1mm]
GW151226~& $20.5^{+6.4}_{-1.5}$ & $0.74^{+0.07}_{-0.05}$\\
\addlinespace[1mm]
GW170104~& $49.1^{+5.2}_{-3.9}$ & $0.66^{+0.08}_{-0.10}$\\
\addlinespace[1mm]
GW170608~& $17.8^{+3.2}_{-0.7}$ & $0.69^{+0.04}_{-0.04}$\\
\addlinespace[1mm]
GW170729~& $80.3^{+14.6}_{-10.2}$ & $0.81^{+0.07}_{-0.13}$\\
\addlinespace[1mm]
GW170809~& $56.4^{+5.2}_{-3.7}$ & $0.70^{+0.08}_{-0.09}$\\
\addlinespace[1mm]
GW170814~& $53.4^{+3.2}_{-2.4}$ & $0.72^{+0.07}_{-0.05}$\\
\addlinespace[1mm]
GW170818~& $59.8^{+4.8}_{-3.8}$ & $0.67^{+0.07}_{-0.08}$\\
\addlinespace[1mm]
GW170823~& $65.6^{+9.4}_{-6.6}$ & $0.71^{+0.08}_{-0.10}$\\
\addlinespace[1mm]
\end{tabular}
\caption{Mass and dimensionless spin of the remnants of binary BH merger candidates observed by the LIGO/Virgo collaboration in the O1 and O2 runs~\cite{LIGOScientific:2018mvr}.}
\label{tab:ligofactsheets}
\end{table}

\section{Cloud formation and gravitational wave emission}\label{sec:setup}

In what follows we assume a scenario in which a bosonic condensate forms around the post-merger remnant. We assume that the colliding BHs are not surrounded by a boson cloud prior to merger. 
If inspiralling BHs are surrounded by a cloud, level transitions can reduce the size of the cloud and, in some cases, deplete it well before merger~\cite{Baumann:2018vus}. 
%
Therefore, a full numerical evolution of a binary BH system with a cloud surrounding one or both BHs is necessary to determine the final state of the cloud(s).

The overall dynamics of the BH/boson cloud system can be described in terms of two competing processes: the growth of the cloud due to the superradiant instability, and its dissipation due to GW emission. These processes can be considered independently, because they act on very different timescales. The cloud grows on the superradiant instability timescale
\be
\tau_{\textrm{inst}} \sim 10 \; \textrm{yr}\,(M_2^8\mu_{13}^9\chi)^{-1},
\ee
where $M_2=M/(10^2\msun)$ and $\mu_{13}=m_s/(10^{-13})$eV.
GW emission, on the other hand, occurs on the timescale
\be 
\tau_{\textrm{GW}} \sim 5\times 10^{7}\; \textrm{yr}\,(M_2^{14}\mu^{15}_{13}\chi)^{-1}.
\ee
These simple analytic expressions for the two timescales can be derived in the limit $M\mu \ll 1$, but they are a fairly good approximation even when $M\mu \sim 1$: cf. Fig.~4 of~\cite{Brito:2017zvb}. Since $\tau_{\textrm{GW}} \gg \tau_{\textrm{inst}}$, we can approximate the system to be a cloud that forms quasi-adiabatically on a fixed Kerr background and use BH perturbation theory to estimate the emitted radiation. Backreaction effects can be neglected since the mass of the cloud is a small fraction of the BH mass, spread out over a large spatial volume~\cite{Brito:2014wla}. The separation in the instability and GW emission timescales allows us to use equations (25) and (26) of~\cite{Brito:2017zvb} to estimate the mass and spin of the BH merger remnant after the cloud forms around it, which are then used as input to compute the GW signal. 
We adopt the framework employed in \cite{Brito:2017wnc,Brito:2017zvb}, which we briefly review below for the sake of completeness (see also \cite{Isi:2018pzk} for a detailed description of the signal morphology).

We will consider ultralight scalars on a fixed Kerr background that satisfy the Klein--Gordon equation $\nabla_{\mu}\nabla^{\mu}\Psi=\mu^2\Psi$. The time and angular dependence of the field can be separated by writing
\be
\Psi = \mathrm{Re}\bigg[\sum_{\ell,m}\int d\omega \, e^{-i\omega t+i m \varphi}{}_0S_{\ell m \omega}(\vartheta)\psi_{\ell m\omega} (r)\bigg], \label{scalarfield}
\ee
where the functions
${}_sS_{\ell m\omega}(\vartheta)$
are spin-weighted spheroidal harmonics~\cite{Berti:2005gp} of spin $s$, orbital number $\ell$ and azimuthal number $m$, with $\ell \ge |s|$ and $|m| \le \ell$. The radial component $\psi_{\ell m\omega} (r)$, on the other hand, is a confluent-Heun type function that satisfies a Teukoslky-like equation for massive scalar fields of frequency $\omega$~\cite{Brill:1972xj}. In particular, the presence of a mass term in the field equation implies the existence of unstable quasibound states for the boson, which are characterized by a discrete set of complex eigenfrequencies $\omega=\omega_R+i\omega_I$, where $|\omega_R| \le \mu$ and $\omega_I \ge 0$~\cite{Cardoso:2005vk, Dolan:2007mj,Berti:2009kk}.\footnote{We note that in equation (8) of \cite{Brito:2017zvb} the imaginary part of $\omega$ should be multiplied by an overall factor $1/M$.} This instability is driven by superradiance~\cite{1971JETPL..14..180Z,Press:1972zz,1972BAPS...17..472M,staro1,staro2,Brito:2015oca,Torres:2016iee}. If the cloud grows much faster than the duration of the GW signal, the gravitational radiation is predominantly emitted at constant frequency $\tilde\omega=2 \omega_R$ (but see~\cite{DAntonio:2018sff,Isi:2018pzk} for data analysis methods taking into account possible frequency drifts).

We use the Teukolsky formalism~\cite{Teukolsky:1973ha} to compute the GWs emitted by the bosonic cloud that forms around the rotating BH. Gravitational radiation is encoded in the Newman-Penrose scalar $\psi_4=\psi_4(t,r,\vartheta,\varphi)$, given by 
\be
\psi_4 =\rho^4 \int_{-\infty}^{\infty} d \omega\sum_{ \ell  m}R_{\ell  m \omega}(r){}_{-2}S_{ \ell m  \omega}(\vartheta)e^{i m \varphi-i \omega t},
\ee
where $\rho=(r-ia\cos\vartheta)^{-1}$ and $R_{\ell  m  \omega}(r)$ satisfies the Teukolsky equation for spin-$2$ fields sourced by the stress energy tensor of the scalar field~\cite{Teukolsky:1973ha,Sasaki:2003xr}. 
 \begin{figure}[t!]
\includegraphics[scale=1.07]{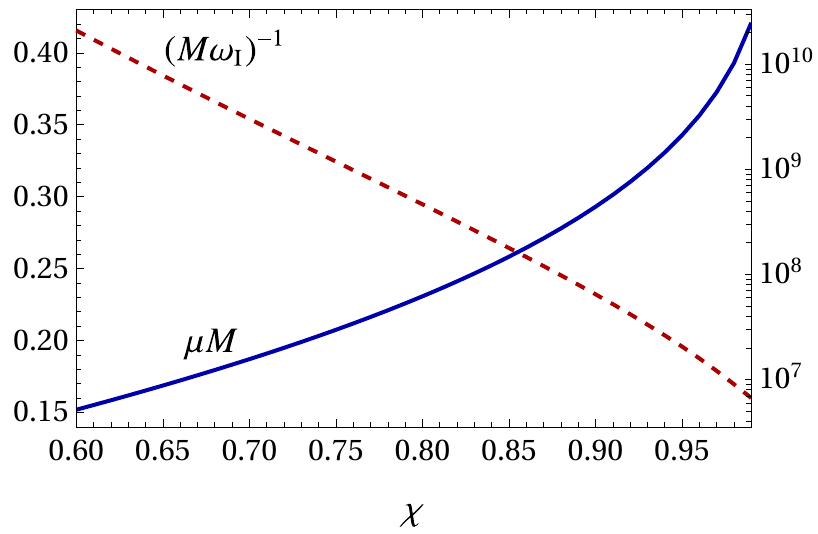}
\caption{The value of $M\mu$ (left $y$-axis) maximizing the  superradiant instability timescale $1/(M\omega_I)$ (right $y$-axis)  of the dominant scalar field mode ($\ell=m=1$) as a function of the  dimensionless BH spin $\chi$. The superradiant instability timescale on the right $y$-axis is in natural units. To convert it to seconds, it should be multiplied by $GM/c^3$ (for quick estimates, note that  $GM_{\odot}/c^3\approx 5\times 10^{-6}$~s).}
\label{peakmu}
\end{figure} 

At infinity, $\psi_4$ reduces to
\be
\psi_4=\frac{1}{r} \int_{-\infty}^{\infty} d \omega \sum_{\ell m}Z^{\infty}_{ \ell m \omega}{}_{-2}S_{ \ell m  \omega}(\vartheta)e^{i m \varphi + i \omega (r-t)},
\label{psi4infinity}
\ee
where $Z^{\infty}_{\ell m \omega}$ are constants related to the energy flux of outgoing gravitational radiation. The Newman-Penrose scalar at infinity is related to the two GW polarizations $h_+$ and ${h}_{\times}$ by 
\be
\psi_4=\frac{1}{2}(\ddot{h}_+-i\ddot{h}_{\times}),
\label{psi4polar}
\ee
where dots represent time derivatives. From Eqs.~\eqref{psi4infinity} and \eqref{psi4polar} we obtain the GW strain $H \equiv h_{+}-ih_{\times}$ in terms of the coefficients $Z^{\infty}_{\ell m \omega}$:
\be
H=-\frac{2}{r} \int_{-\infty}^{\infty}\frac{ d \omega}{\omega ^2}\sum_{ \ell m}Z_{ \ell  m \omega}^{\infty}{}_{-2}S_{\ell m  \omega}(\vartheta) e^{i 
m \varphi + i \omega(r-t)}\,. \label{Hexp}
\ee
The GW energy flux at infinity is
\be
\dot{E} =  \int d\omega \, d \Omega \, r^2 \, \frac{\dot{h}_+^2+\dot{h}_{\times}^2}{16\pi} =  \int_{-\infty}^{\infty}  d \omega\sum_{\ell m}\frac{|Z^{\infty}_{\ell m \omega}|^2}{4\pi \omega ^2}, \label{dEdt}
\ee
where $d\Omega$ denotes the solid angle element and we have used the angular average $\left<|_sS_{\tilde \ell \tilde m\tilde\omega}|^2\right>=1/(4\pi)$.

Through its stress-energy tensor, each mode $(\omega,\ell,m)$ of the scalar field \eqref{scalarfield} acts as a source of quadrupolar radiation, emitting monochromatic GWs with frequency $\tilde{\omega}= 2\omega$, azimuthal number $\tilde{m}=2 m$, and orbital number $\tilde \ell \ge 2 \ell$~\cite{Yoshino:2013ofa,Brito:2017zvb}. We will conservatively (and for simplicity) assume that only the fastest growing scalar mode with $\ell=m=1$ contributes to the cloud~\cite{Dolan:2007mj}. For reference, in Fig.~\ref{peakmu} we show the value of $M\mu$ corresponding to the maximum instability growth rate as a function of BH spin. 

We can rewrite the GW strain in Eq.~\eqref{Hexp} as  
\beq
  H=-\sum _{\tilde \ell}\frac{2 Z^{\infty}_{\tilde \ell}}{\tilde\omega^2r}
  \left({}_{-2}S_{\tilde \ell\tilde m\tilde \omega}e^{i\left[\tilde \omega(r-t)+ \tilde m \varphi \right]}
  \right.\nonumber\\
  \left.
+_{-2}S_{\tilde \ell-\tilde m-\tilde \omega}e^{-i\left[\tilde \omega(r-t) - \tilde m \varphi \right]}\right)\,,
\eeq
where we wrote $Z^\infty_{\tilde \ell} = Z^{\infty}_{\tilde \ell\tilde{m} \tilde{\omega}}$ for brevity, and we have taken into account the symmetries of the system with respect to the transformation $(\omega,m) \rightarrow (-\omega,-m)$. Since ${}_{s}S_{\ell m \omega}(\theta) \in \mathbb{R}$ when $\omega \in \mathbb{R}$, the GW strains in each polarization mode $h_{+}=\mathrm{Re}(H)$ and $h_{\times}=\mathrm{Im}(H)$ are
\beq\label{strainamplitude}
h_{+} =-\sum_{\tilde \ell}h^{(\tilde \ell)}_0\left\{ {}_{-2}S_{\tilde \ell
    \tilde{m}\tilde{\omega}}\cos\left[\tilde\omega(r-t) +\tilde m\varphi +\phi_{\tilde \ell}\right] \right. \nonumber \\
     + \left. {}_{-2}S_{\tilde \ell
    -\tilde{m}-\tilde{\omega}}\cos\left[\tilde\omega(r-t) +\tilde m\varphi -\phi_{\tilde \ell}\right]\right\}, \ \ \ \ \
    \\
h_{\times}=-\sum_{\tilde \ell}h^{(\tilde \ell)}_0\left\{ {}_{-2}S_{\tilde \ell
    \tilde{m}\tilde{\omega}}\sin\left[\tilde\omega(r-t) +\tilde m\varphi +\phi_{\tilde \ell}\right] \right. \nonumber \\
     + \left. {}_{-2}S_{\tilde \ell
    -\tilde{m}-\tilde{\omega}}\sin\left[\tilde\omega(r-t) +\tilde m\varphi -\phi_{\tilde \ell}\right]\right\}, \ \ \ \ \ 
\eeq
where the angle $\phi_{\tilde \ell}$ is the phase of $Z^{\infty}_{\tilde \ell}$ and, following~\cite{Isi:2018pzk}, we have defined the GW intrinsic strain amplitude:
\be\label{amplitude}
h^{(\tilde \ell)}_0=\frac{2|Z^{\infty}_{\tilde \ell}|}{\tilde\omega^2r}\,.
\ee
Finally, the GW strain measured at the detector is
\be
h=h_+F_++h_{\times}F_{\times}\,,
\ee
where $F_{+,\times}$ are antenna pattern functions that depend on the orientation of the detector and the direction of the source (see e.g.~\cite{Thorne:1987af}
for explicit expressions). In our estimates we will only consider the gravitational mode $\tilde \ell=2$, which is the dominant GW mode in the parameter space of interest (see Appendix~\ref{app:modes}).

\begin{figure}[t!]
\includegraphics[scale=0.65]{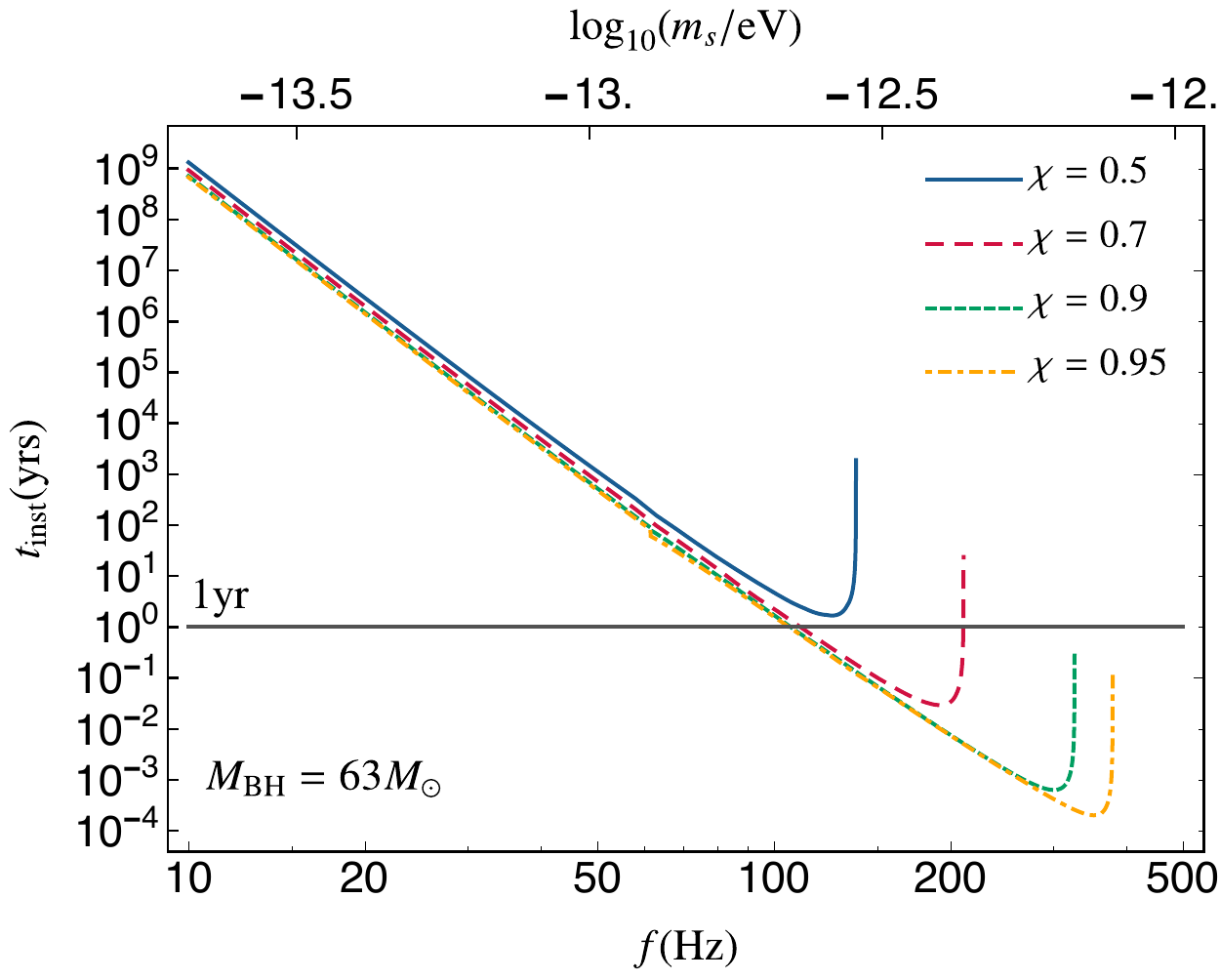}
\caption{Instability timescale for a remnant BH of mass $63 \, \msun$ (the GW150914 remnant mass) and selected values of the remnant BH spins. The timescale (in years) is plotted as a function of both the GW frequency (bottom \textit{x}-axis) and of the boson mass in eV (top \textit{x}-axis).}
\label{fig:GW150914}
\end{figure}

The superradiant instability must develop fast enough for the GW signal to be detectable within the observation time of a given GW detector. Figure~\ref{fig:GW150914} shows the typical instability timescales associated with cloud growth around a BH of mass $63~M_\odot$ (the GW150914 remnant mass~\cite{Abbott:2016blz}) for selected values of the BH spins. The cloud formation timescales are plotted as a function of the scalar field mass (top $x$-axis) and as a function of the corresponding GW frequency (bottom $x$-axis).  For a GW150914-like system, the cloud can grow in less than one year only if the remnant spin $\chi\gtrsim 0.5$.

%
\section{Detection prospects and constraints on the boson mass}\label{sec:constraints}

 %
\begin{figure}[t!]
\includegraphics[scale=0.39]{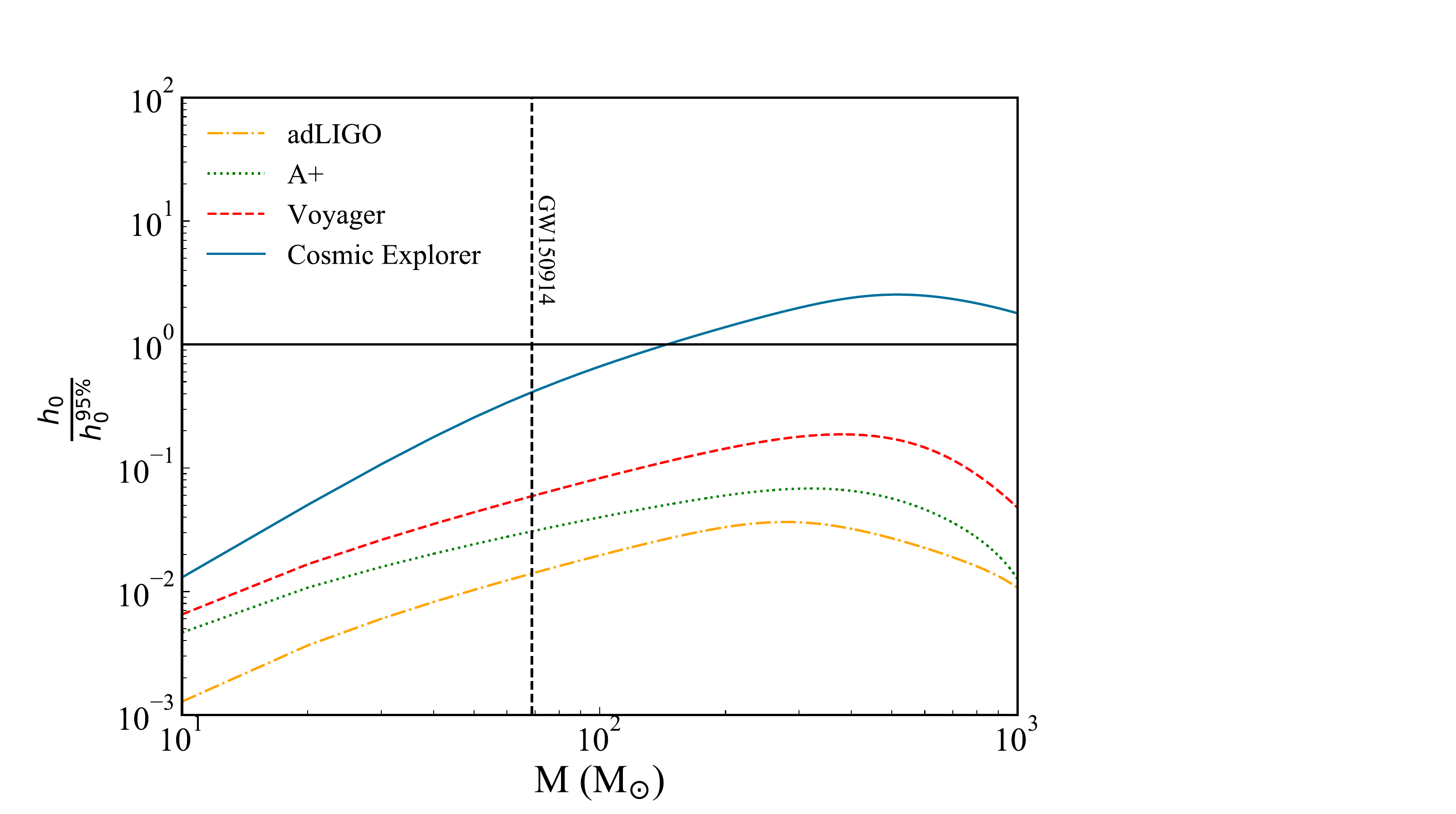} 
\caption{Relative intrinsic amplitude above the 95\%-confidence strain upper limit at AdLIGO, A$+$, Voyager and Cosmic Explorer for GW signals from BHs with spin $\chi= 0.7$ at redshift $z=0.1$ and detector-frame mass in the range $[10,1000]  \, \msun$. We choose the boson mass such that $M \mu$ maximizes the intrinsic amplitude, and we compute $h_0^{95\%}$ using a coherent observation time $T_{\textrm{coh}}=10$~days and a number of segments such that the total observation time is 2 years.}
\label{fig:spin07}
\end{figure}

\begin{figure*}[htb]
\begin{tabular}{cc}
\includegraphics[scale=0.415]{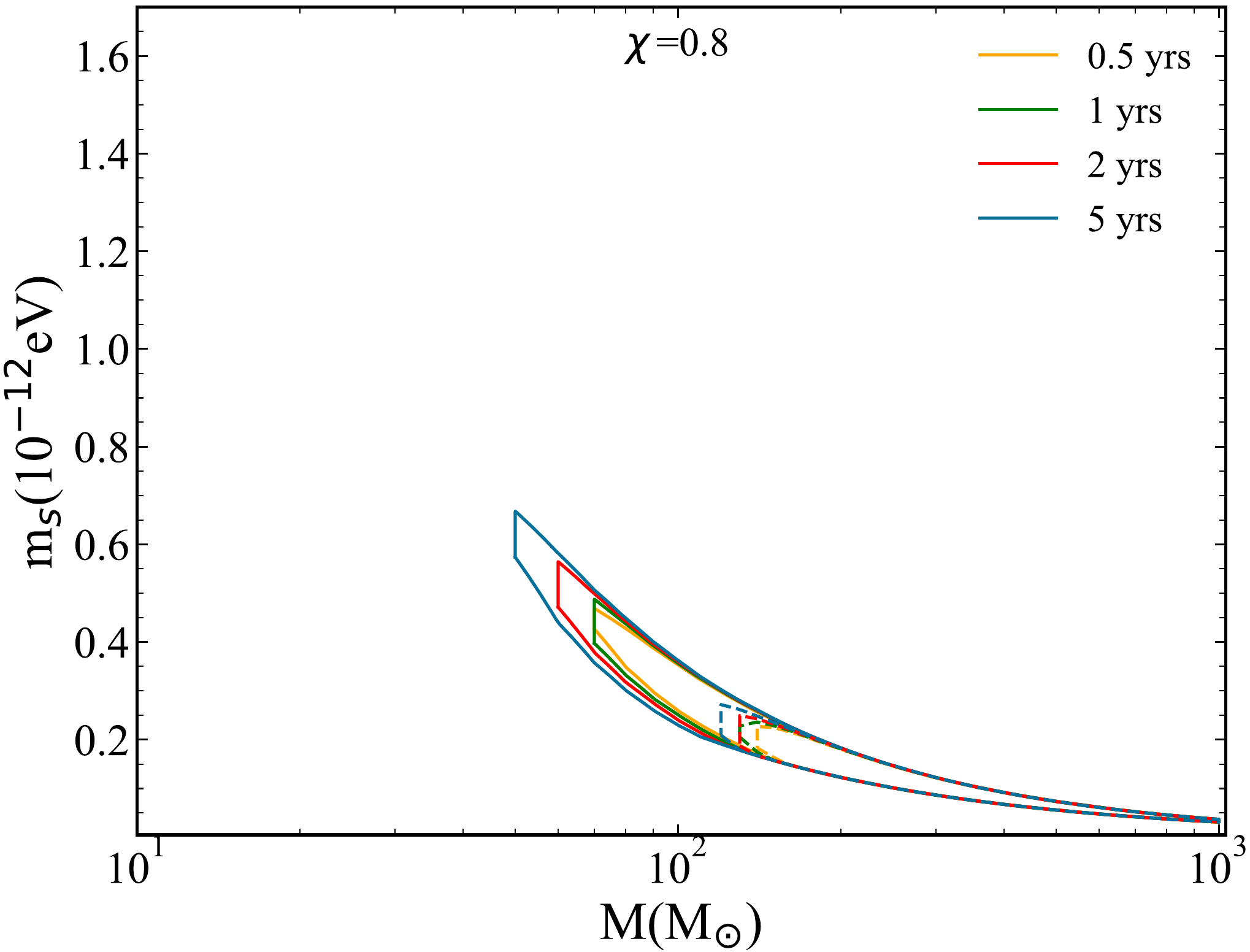}
\includegraphics[scale=0.415]{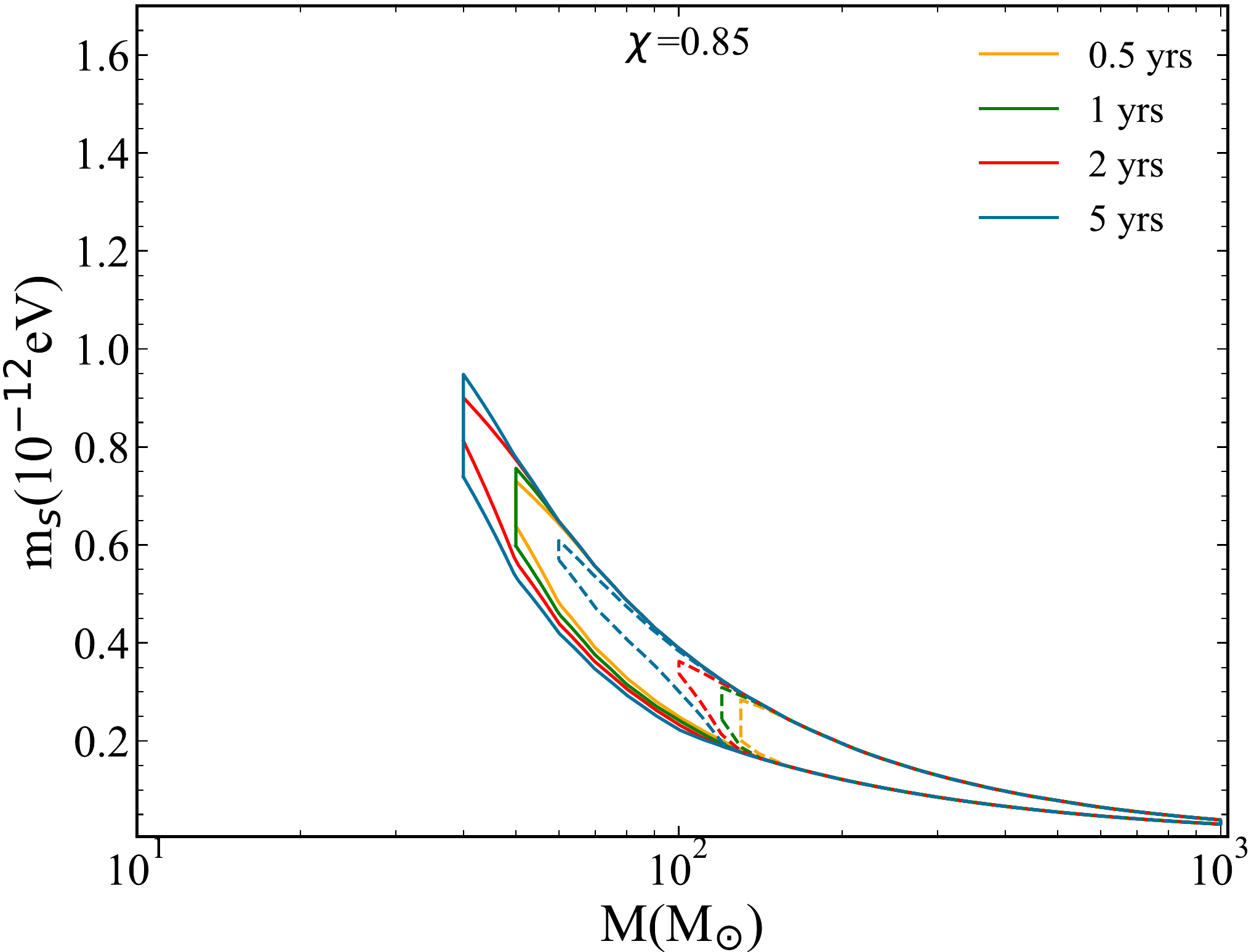}  \\
\includegraphics[scale=0.415]{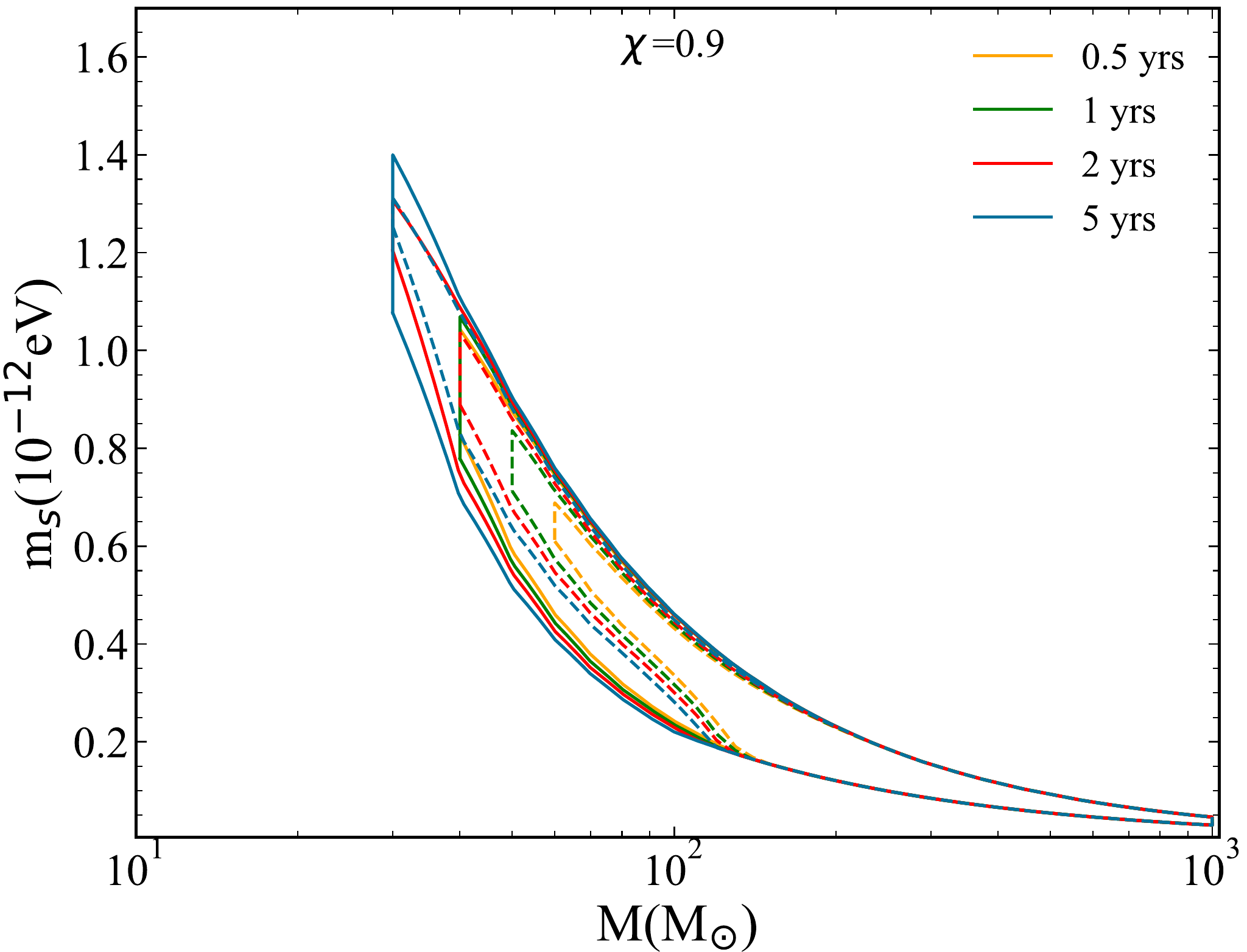} 
\includegraphics[scale=0.415]{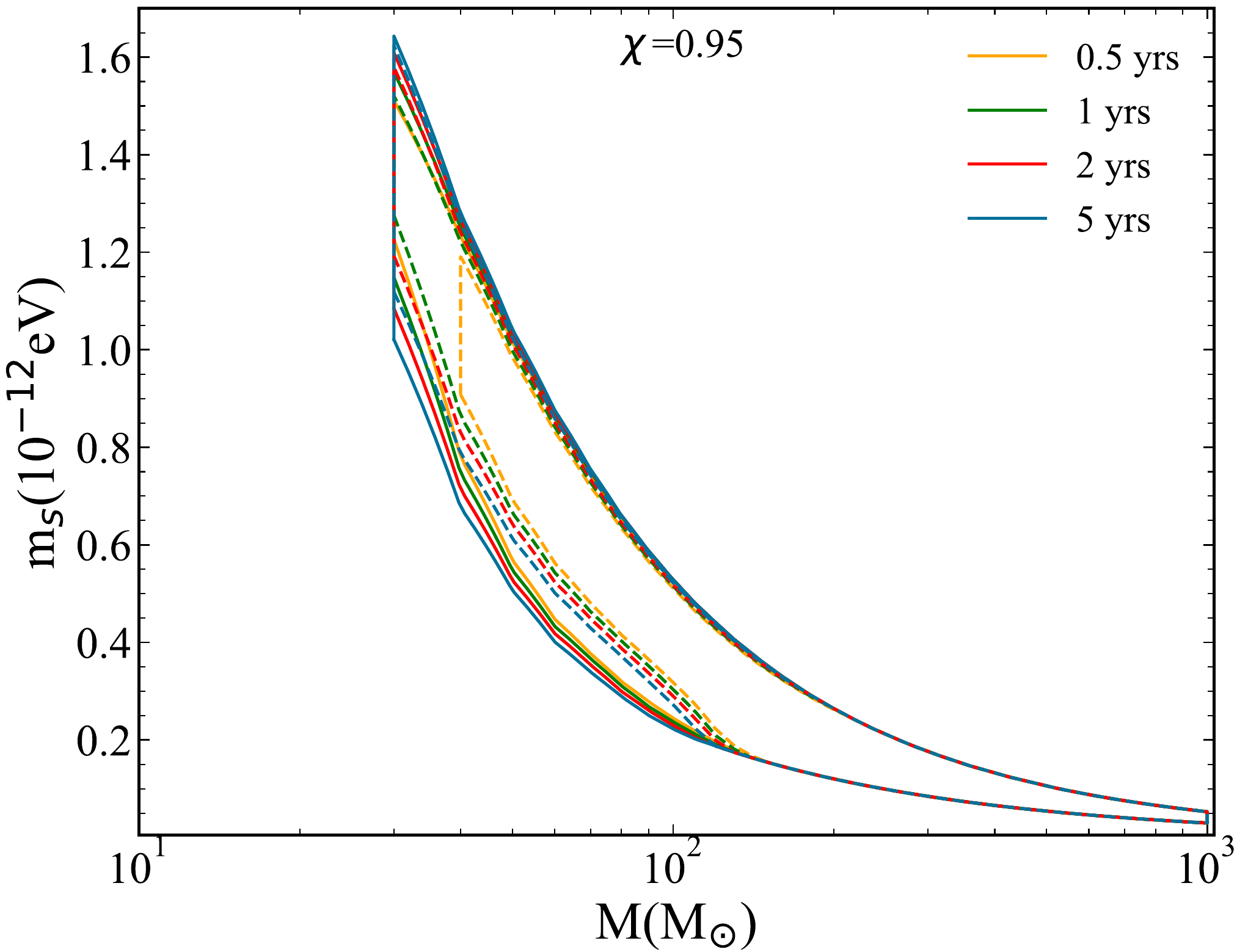} 
\end{tabular}
\caption{Contour plot in the $(M,\,m_s)$ plane (where $M$ is the detector-frame BH mass) of BH/cloud signals at $z=0.1$ that would detectable by Cosmic Explorer. We consider a semi-coherent search with $T_{\rm coh}=10$ days and different observation times (indicated by the different colors) with (\textit{dashed}) and without (\textit{solid}) self-confusion noise. Our estimate of the self-confusion noise is described in the main text. For all plots we consider superradiant instabilities that grow within $30$ days. The four panels correspond to different BH spins $\chi$, as indicated in the legend.}
\label{fig:massrange}
\end{figure*}

Having laid down our framework to compute the GW signal from a BH/boson cloud system, let us now assess the parameter space for which these signals would be detectable by present and future detectors, including AdLIGO, A$+$, Voyager and Cosmic Explorer. 

For long-lived signals, such as the ones we are interested in, it is usually preferable to use a semi-coherent search method. In particular, when the source parameters are uncertain or not known \textit{a priori}, the search spans a broad frequency band which makes a fully coherent search computationally prohibitive. In semi-coherent search methods, the signal is first divided into $N$ segments of fixed length $T_{\textrm{coh}}$, such that the total observation time is $N \times T_{\textrm{coh}}$. 
Semi-coherent search methods specifically aimed at BH/cloud systems have recently been proposed~\cite{DAntonio:2018sff,Isi:2018pzk}. In particular, Ref.~\cite{Isi:2018pzk} estimated the upper limit on the GW strain amplitude necessary for detection. For signals with well-known sky locations, but unknown inclination and polarization angle, the $95\%$-confidence strain upper limit is roughly given by
\begin{equation}
h_0^{95\%}\approx\frac{25}{N^{1/4}}\sqrt{\frac{S_h(f)}{N_{\rm ifo} T_{\textrm{coh}}}}\,,
\label{eq:threshold}
\end{equation}
for a network of $N_{\rm ifo}$ detectors with comparable power spectral density (PSD)  $S_h(f)$ at the signal frequency. We conservatively assume $N_{\rm ifo} =1$. As done in~\cite{Isi:2018pzk} we will consider boson signals detectable if they reach an intrinsic amplitude of $h_0^{95\%}$ or higher. Ideally, the coherent time of integration $T_{\rm coh}$ would be  chosen case-by-case such that, over a time $T_{\rm coh}$, the frequency varies by at most $\sim 1/T_{\rm coh}$~\cite{Isi:2018pzk}. However, the frequency drift of these signals -- due to the cloud's self-gravity and possible axion self-interaction -- is poorly known, although it is expected to be very small for most of the parameter space~\cite{Arvanitaki:2014wva,Isi:2018pzk}. To facilitate comparisons with previous studies~\cite{Arvanitaki:2016qwi} we set $T_{\rm coh}=10$ days.

In the first two observing runs, the LIGO/Virgo collaboration detected ten binary BH merger candidates~\cite{LIGOScientific:2018mvr}. The masses and spins of the merger remnants are listed in Table~\ref{tab:ligofactsheets}. The merger of intermediate-mass BHs is also an important target for LIGO/Virgo searches~\cite{Abbott:2017iws,CalderonBustillo:2017skv} and, as shown in~\cite{Isi:2018pzk}, the most promising source of continuous post-merger GWs from BHs surrounded by bosonic clouds. Therefore we consider merger remnants with masses in the range $[10,\,1000]~\msun$. The upper limit in BH mass is motivated by the low-frequency cutoff of the detectors, that we assume to be at $10$ Hz. In this range of BH masses we can probe about two orders of magnitude in boson masses, as we show below.

All merger remnants observed by AdLIGO so far have dimensionless spins in the range $0.66 \lesssim \chi \lesssim 0.81$ and source-frame masses in the range $17.8\msun\lesssim M\lesssim 80.3\msun$, as seen in Table~\ref{tab:ligofactsheets}. The superradiant instability grows at the expense of the rotational energy of the BH, so higher masses and higher BH spins (as in the recently announced GW170729) favor the growth of a cloud and yield a larger GW amplitude.

For illustration, in Figure~\ref{fig:spin07} we show that for $\chi=0.7$, none of the current and planned detectors could detect the signal from the cloud for a GW150914-like event\footnote{Despite the higher mass and spin of the remnant of GW170729, its follow-up signals have a lower signal-to-noise ratio compared to those of GW150914, due to its much larger redshift.} with $M\lesssim 70 \, \msun$ at $z=0.1$ (or luminosity distance $D_L\sim 475$~Mpc) in two years of observation with $T_{\textrm{coh}}=10$~days. Cosmic Explorer could detect signals from remnants with spin $\chi\sim 0.7$ and redshift $z=0.1$, but only for source-frame masses $\gtrsim 150\msun$. These results are consistent with Ref.~\cite{Isi:2018pzk}.

With these considerations in mind, we look more closely at the parameter space that could be probed by Cosmic Explorer. In Fig.~\ref{fig:massrange} we consider Cosmic Explorer sources at redshift $z=0.1$ with selected dimensionless birth spins ($\chi = 0.8, 0.85, 0.9, 0.95$), and we show contours corresponding to the region where the signal would be detectable in the $(M,\,m_s)$ plane, where $M$ is the detector-frame BH mass. For any $M$, the upper limit on $m_s$ (or $\mu$) corresponds to the value of $M\mu$ at which the instability cuts off; the lower bound corresponds to an instability growth time of 30 days.\footnote{The limits would not change significantly had we chosen larger values for the instability timescale, because the strain amplitude drops very rapidly for small values of $M\mu$~\cite{Isi:2018pzk}.}

This plot can be interpreted in two ways: once a binary BH merger is observed and the remnant mass $M$ is known, one can read off from this plot the minimum observation time that would be required to detect a superradiant signal for boson masses in the range shown in the figure. Alternatively, in the absence of a detection, the plot shows the range of boson masses that can be ruled out.

\begin{figure}[thb]
\includegraphics[scale=0.39]{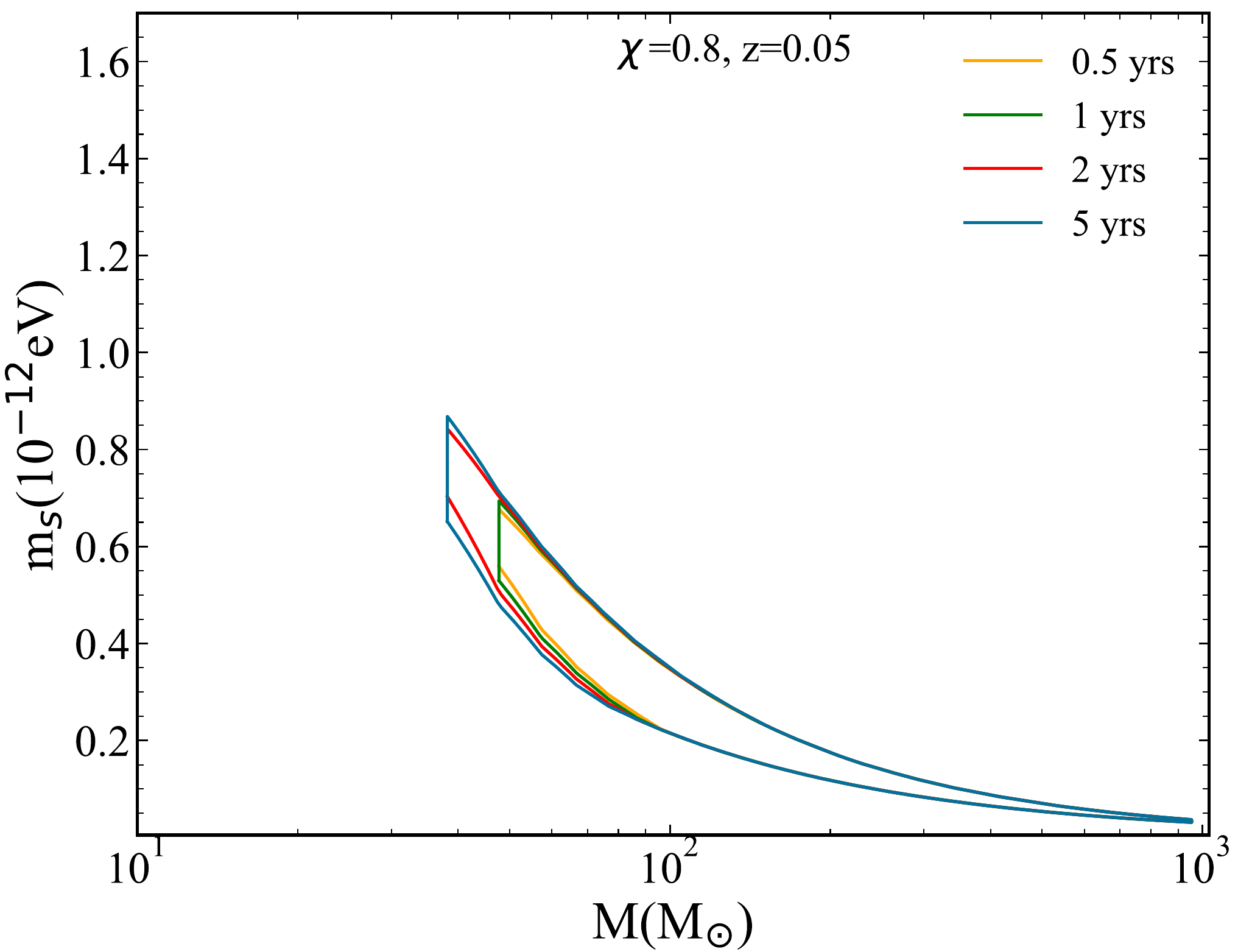}
\caption{Same as Fig.~\ref{fig:massrange} for a system with BH spin $\chi=0.8$ detected at smaller redshift ($z=0.05$).}
\label{fig:massrangez0p05}
\end{figure}

The strain amplitude is obviously higher for sources at a lower redshift, hence a larger part of the parameter space would be detectable for the same observation time. For illustration, in Fig.~\ref{fig:massrangez0p05} we show how the contour plot would change for events with $\chi=0.8$ (corresponding to the top-left panel of Fig.~\ref{fig:massrange}) detected at redshift $z=0.05$.

\begin{figure}[tb]
\includegraphics[scale=0.39]{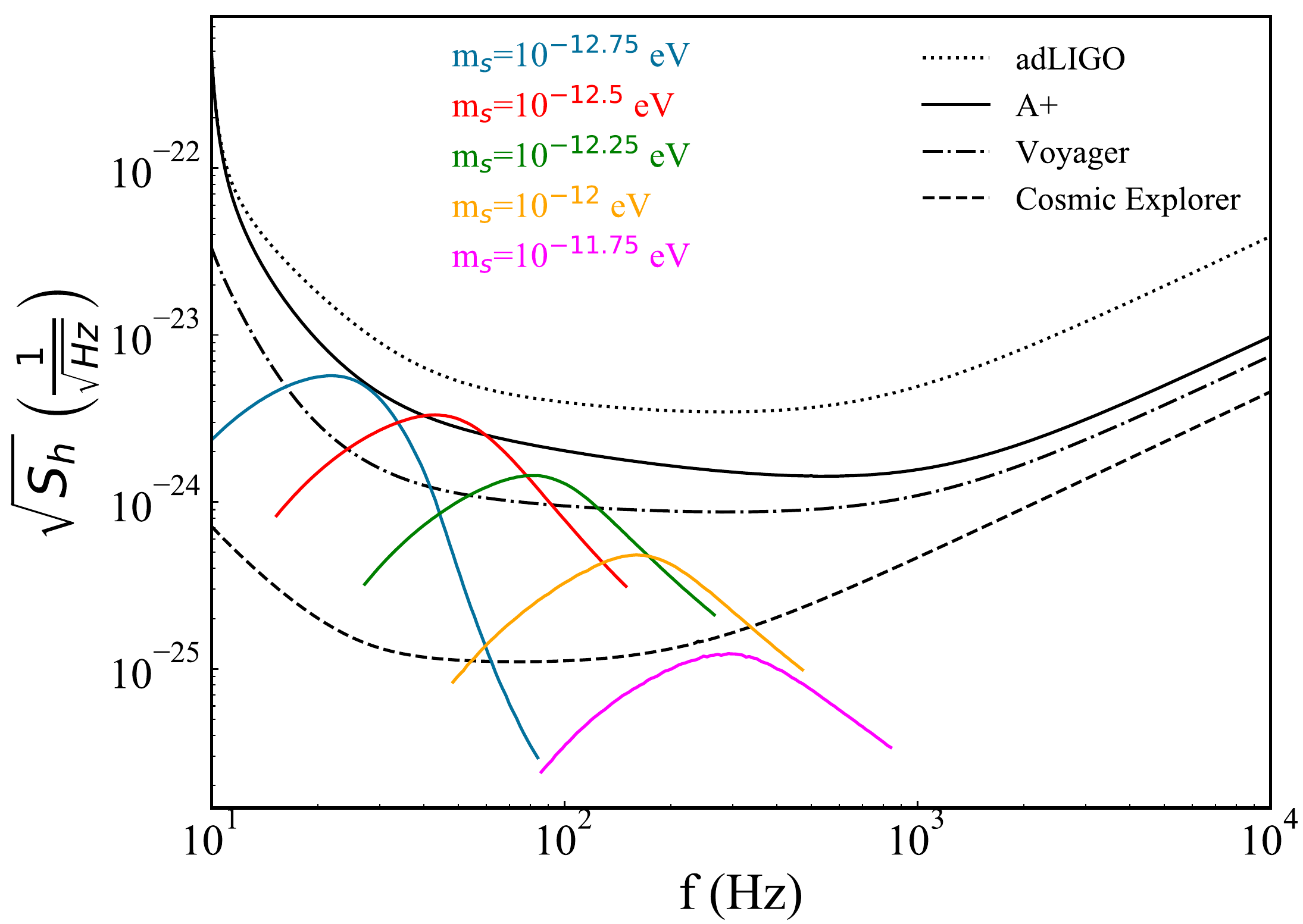}
\caption{Stochastic background fluxes (in color) plotted over the noise PSD (in black) for the different detectors.}
\label{fig:PSD}
\end{figure}

\subsection{Self-confusion noise from a stochastic background}

So far we have considered instrumental noise as the only noise source. However, if light scalar fields exist,
signals which are too faint to be individually resolved could contribute to a {\em stochastic background}, which could be strong enough to be a source of confusion noise (especially for future detectors)~\cite{Brito:2017wnc}.

Here we follow~\cite{Brito:2017wnc} to estimate this stochastic background. Most of the background is produced by isolated BHs, therefore we neglect the contribution from binary BH mergers. We also neglect the astrophysical background from other sources, such compact binaries or continuous waves from neutron stars,  since those are expected to be much smaller than the background from boson clouds in the relevant frequency range: compare Fig.~6 of~\cite{Regimbau:2011rp} with Fig.~2 of~\cite{Brito:2017wnc}.
 
 To be conservative, we maximize the background contribution by adopting the most optimistic spin magnitude distribution of~\cite{Brito:2017wnc} (i.e., we assume that spin magnitudes are uniformly distributed in the range $\chi\in [0.8,\,1]$). The background's energy spectrum can be translated to a spectral density -- cf. Eq.(3) of~\cite{Thrane:2013oya} -- that we add to the instrumental PSD in order to estimate the full noise, in analogy with the familiar galactic white dwarf binaries confusion noise in the LISA band~\cite{Cornish:2017vip}.

As shown in Fig.~\ref{fig:PSD}, the confusion noise background effectively increases the noise PSD of the detector within a frequency range that depends on the boson mass. The background is not expected to have a large impact on the noise budget for AdLIGO and A$+$, but it could become significant for Voyager and Cosmic Explorer at frequencies $\sim 30$--$300$ Hz, corresponding to bosons in the mass range $\sim10^{-12}$--$10^{-12.75}$ eV.

This additional noise source can significantly impact our ability to perform follow-up searches from BH/cloud remnants. By comparing the solid and dashed lines in Fig.~\ref{fig:massrange}, we see that the background could significantly lower detection prospects for clouds that form around BHs with detector-frame masses $M\lesssim 100\msun$ and spins $\chi\lesssim 0.85$.

In Table~\ref{tab:detectorsrange}  we list the range of detectable boson masses for AdLIGO, A$+$ and Voyager for boson clouds that grow within 30 days. at redshift $z = 0.1$. For this table we assume an observation time of 2 years and $T_{\rm coh}= 10$~days. For each detector and for each value of $\chi$ we list the lowest BH mass that would be detectable and the corresponding boson mass range. For AdLIGO and A+ there is also an upper limit on the detectable BH mass ($770 \msun$ for AdLIGO when $\chi=0.95$, and $770 \msun$ for A$+$ when $\chi=0.9$). This is due to the lower sensitivity of these detectors at frequencies $\sim 10$~Hz. Table~\ref{tab:detectorsrange} also illustrates the effect of the stochastic background. As expected, the confusion noise hardly affects the results for AdLIGO. For both AdLIGO and A$+$, the only sources that can be detected are clouds that form around highly-spinning intermediate-mass BHs. Voyager may be able to detect superradiant instabilities of BHs with masses $M \lesssim 100 \msun$, but only if they are highly spinning. These results are consistent with Ref.~\cite{Isi:2018pzk}.

\begin{table*}
\begin{center}
\begin{tabular}{ccccccc}
\hline
\hline
\multirow{2}{*}{Detector} & \multicolumn{2}{c}{$\chi=0.85$} & \multicolumn{2}{c}{$\chi=0.9$} & \multicolumn{2}{c}{$\chi=0.95$} \\
\cmidrule{2-7}
 & M ($\msun$) & m$_s$ ($10^{-13}$~eV) & M ($\msun$) & m$_s$ ($10^{-13}$~eV) & M ($\msun$) & m$_s$ ($10^{-13}$~eV) \\
\hline
\hline
\multirow{2}{*}{AdLIGO}  & -- & -- & -- & -- & 280 & 1.50 -- 1.59 \\
 & -- & -- & -- & -- & (280) & (1.51 -- 1.58)  \\\\
\multirow{2}{*}{A$+$}  & -- & -- & 260 & 1.49 -- 1.53  & 110 & 3.82 -- 4.04 \\
 & -- & -- & (260) & (1.50 -- 1.51) & (140) & (2.99 -- 3.16)  \\\\
\multirow{2}{*}{Voyager}  & 200 & 1.72 -- 1.82 & 100 & 3.70 -- 4.04 & 60 & 6.53 -- 7.64 \\
 & (210) & (1.62 -- 1.74) & (140) & (2.58 -- 2.88) & (70) & (6.00 -- 6.35)  \\
\end{tabular}
\caption{Range of boson masses that can be probed by AdLIGO, A$+$ and Voyager for boson clouds that grow within 30 days at redshift $z=0.1$, assuming a semi-coherent search with $T_{\rm coh}=10$ days and an observation time of 2~years. Only spins $\chi\gtrsim 0.8$ lead to detectable signals. BH masses are solar mass units, while boson masses are in units of $10^{-13}$~eV. Values in parentheses include self-confusion noise, which is estimated as described in the main text.}
\label{tab:detectorsrange}
\end{center}
\end{table*}

\section{Detection rates}\label{sec:rates}

How many binary BH merger remnants could emit observable post-merger GW signals due to the growth of a boson cloud? Under optimistic assumptions and using analytical approximations to the GW amplitude, Ref.~\cite{Arvanitaki:2016qwi} estimated that Cosmic Explorer could see $\gtrsim 100$ such events from scalar clouds (see also Ref.~\cite{Baryakhtar:2017ngi} for similar estimates for vector fields). Here we revisit those estimates using numerical calculations of the GW strain and exploring the impact of astrophysical assumptions on the BH mass and spin distributions. We focus on Cosmic Explorer, since (as shown above and in~\cite{Arvanitaki:2016qwi}) detection prospects for AdLIGO, A$+$ and Voyager are not very promising.
 
We estimate the number of merger events that emit detectable long-lived GWs using~\cite{Hartwig:2016nde}
\begin{equation}\label{RATES1}
{N}=  T_{\rm obs} \int_{h_0 > h_0^{95\%}}4\pi c\frac{d^2 \dot{n}}{d M  d \chi} \frac{dt}{dz} D_c^2 dz d M d \chi \,,
\end{equation}
where 
\begin{equation}
\frac{{\rm} d t}{{\rm} d z}=\frac{1}{H_0 \sqrt{\Delta} (1+z)}\,,
\end{equation} 
is the derivative of the lookback time with respect to redshift,  $D_c$ is the comoving distance and $H_0$ is the local Hubble constant. We also defined $\Delta(z) = \Omega_M(1+z)^3 + \Omega_\Lambda$ where $\Omega_M$ is the dimensionless matter density and $\Omega_{\Lambda}$ is the dimensionless cosmological constant density. For the cosmological parameters we use the latest Planck measurements~\cite{Aghanim:2018eyx}, and we assume a spatially flat Universe.

To compute ${d^2 \dot{n}}/{d M d \chi}$ we assume that the merger rate is independent of the BH mass and spin, such that ${d^2 \dot{n}}/{d M d \chi}=\mathcal{R}(z) P(M) P(\chi)$, with $\mathcal{R}(z)$ the total merger rate per comoving volume per year and $P(M)$, $P(\chi)$ represent the distribution of the remnant's source frame mass and spin, respectively. For the  total comoving merger rate we use the estimates of Ref.~\cite{Dvorkin:2016wac}\footnote{The majority of the mergers that produce a detectable signal have redshifts $z\lesssim 1$. Up to redshift $z=1$ the total comoving merger rate computed in~\cite{Dvorkin:2016wac} is nearly independent on the specific astrophysical model and is very well described by $\mathcal{R}(z)\propto (1+z)^{1.4}$.} normalized to $\mathcal{R}(0)=50\,{\rm Gpc}^{-3}\,{\rm yr}^{-1}$, according to LIGO and Virgo's observed local rate~\cite{LIGOScientific:2018jsj}.\footnote{The $90\%$ credible interval for the local merger rate measured after the first and second observing runs of Advanced LIGO and Advanced Virgo is $52.9^{+55.6}_{-27}\,{\rm Gpc}^{-3}\,{\rm yr}^{-1}$~\cite{LIGOScientific:2018jsj}.}

For the distribution of the progenitor's source-frame masses, $m_1$ and $m_2$, with $m_1>m_2$, we adopt two different prescriptions:

\begin{itemize}

\item[{\bf (i)}] Following the LIGO-Virgo Scientific Collaboration~\cite{TheLIGOScientific:2016pea}, we use a power-law distribution $P(m_1)\propto m_1^{\alpha}\theta(m_1-5\,M_{\odot})$, where $\theta$ represents the Heaviside step function. We use a spectral index $\alpha = -2.35$ for the primary BH, while the secondary mass is uniformly distributed in $m_2\in  [5\,M_{\odot}, m_1]$. For this distribution we impose an upper mass limit $m_1<50\,M_{\odot}$. For short we call this model the power-law  or ``PL'' model.

\item[{\bf (ii)}] We also use a distribution for the primary component given by $P(m_1)\propto m_1^{\alpha}e^{-m_1/m_{\rm cap}}\theta(m_1-5\,M_{\odot})$, with $m_{\rm cap}=60\,M_{\odot}$~\cite{Kovetz:2016kpi}, and the the secondary mass is uniformly distributed in $m_2\in  [5\,M_{\odot}, m_1]$. For this distribution we do not impose an upper mass limit, therefore this distribution allows to study the impact of the possible existence of remnants with masses above $100 M_{\odot}$. For short we call this model the exponentially suppressed or ``ES'' model.

\end{itemize}

The $50\,M_{\odot}$ upper BH mass limit that we impose in the first model is consistent with LIGO's observations and it excludes the detection of remnant BHs with masses above $100\msun$. This distribution is realistic since pair instability and pulsation pair instability in massive helium cores may inhibit the formation of BHs with masses above $\sim 50\,M_{\odot}$~\cite{Belczynski:2016jno}. However, progenitors formed through previous BH mergers can have masses above $50\,M_{\odot}$~\cite{Gerosa:2017kvu,Fishbach:2017dwv}. Mergers involving second-generation BHs could occur in dense stellar environments~\cite{Rodriguez:2016kxx,Samsing:2018isx,DOrazio:2018jnv}. BHs with masses above $200\msun$ can have a Population III origin, but merger rates for those BHs are expected to be very small in the local Universe~\cite{Hartwig:2016nde,Belczynski:2016ieo}.

We assume the dimensionless spin magnitudes to be uniformly distributed in the range $[0, 1]$ for both BHs, and we consider two different prescriptions for their orientations: 

\begin{itemize}

\item[{\bf (i)}] ``Isotropic'' model: The spin directions are isotropically distributed, as expected for BH binaries produced in dense stellar environments~\cite{Rodriguez:2016vmx}. This case tends to produce remnant BHs with spin magnitudes around $\chi\sim 0.7$~\cite{Berti:2008af,Gerosa:2017kvu,Fishbach:2017dwv} and is therefore somewhat pessimistic.

\item[{\bf (ii)}] ``Aligned'' model: We assume the spins to be aligned with the orbital angular momentum, as typically expected for field binaries (see~\cite{Gerosa:2018wbw} for a comprehensive study of spin orientations in this scenario). This model tends to produce more BHs with spin  $\chi\gtrsim 0.7$, and is therefore more optimistic.

\end{itemize}

\begin{figure*}[htb]
\begin{tabular}{cc}
\includegraphics[scale=0.65]{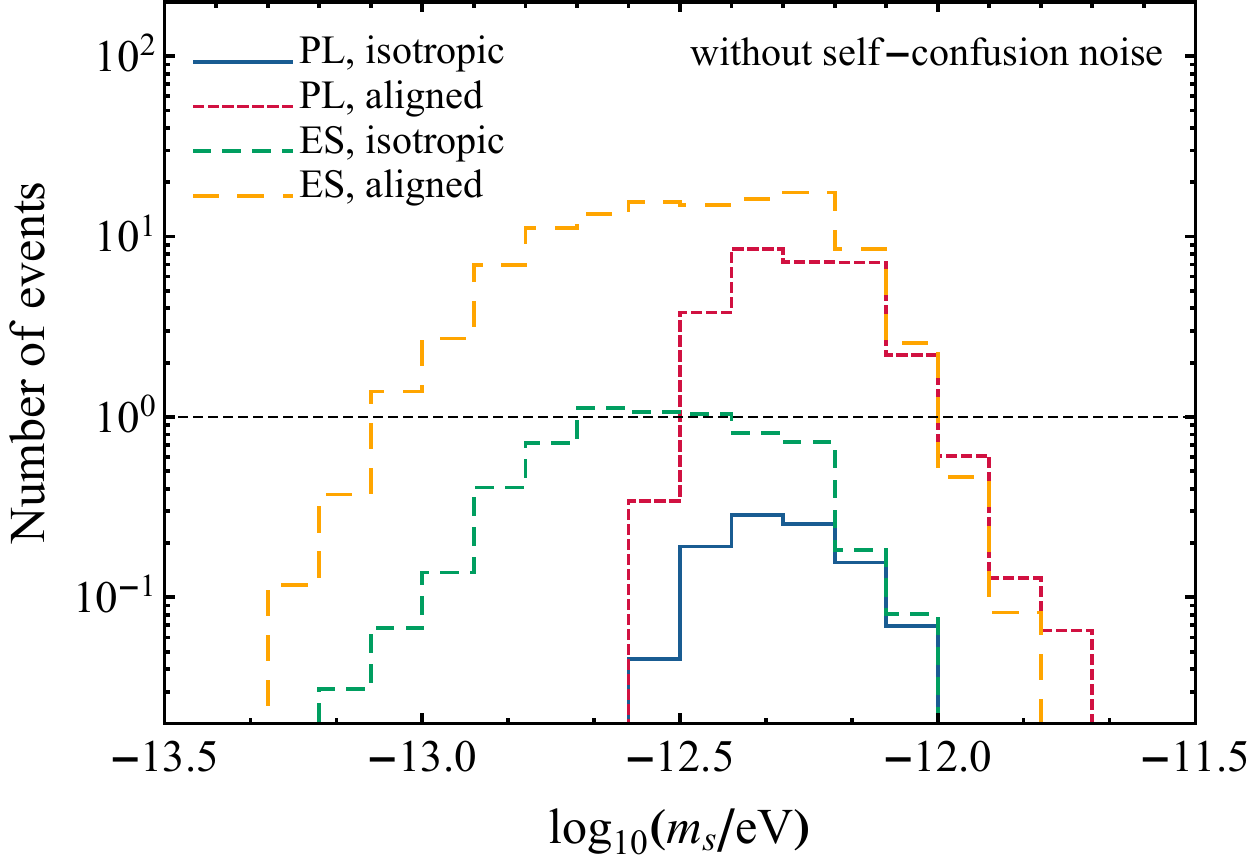}
\includegraphics[scale=0.65]{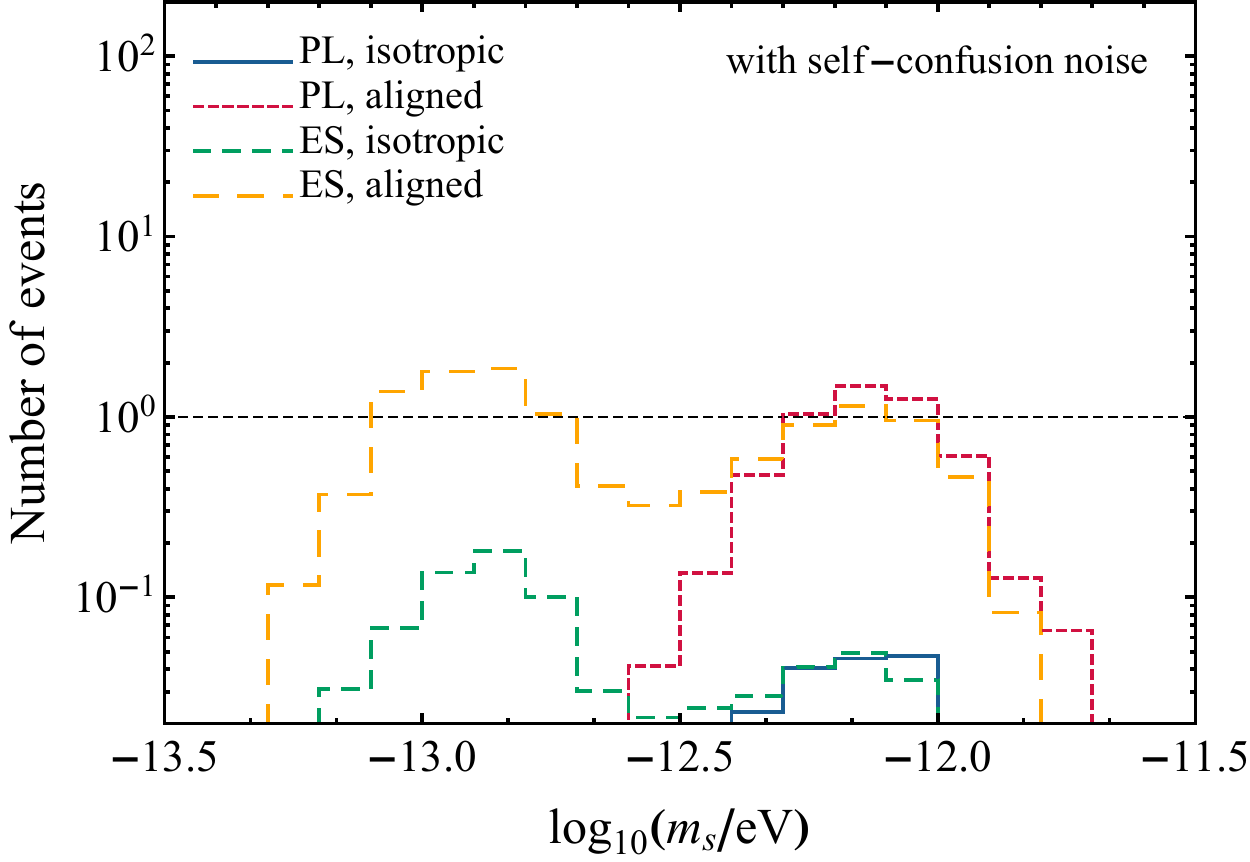}
\end{tabular}
\caption{Merger events that could have detectable post-merger GW signals without (left panel) and with (right panel) self-confusion noise for Cosmic Explorer, assuming $T_{\rm coh}=10$ days and one year of continuous observation time.  The different curves correspond to different astrophysical assumptions on the progenitor masses and spins, as described in the main text. The dashed black line marks the threshold to have at least one observable event within one year of observation.}
\label{fig:rates}
\end{figure*}

To estimate the number of detectable events, we first draw the progenitor properties as outlined above, and then we compute the distributions of the mass and spin of the merger remnant using numerical relativity fitting formulas~\cite{Barausse:2009uz,Barausse:2012qz}. The number of events can then be obtained using Eq.~\eqref{RATES1}. Our results for a coherent integration time of $T_{\rm coh}=10$~days and one year of continuous observation time are shown in Fig.~\ref{fig:rates}.
Detection prospects for follow-up searches are considerably better for the aligned-spin distribution, because larger remnant spins generate post-merger GWs with larger intrinsic strain amplitudes. In addition, due to the $\sim 100M_{\odot}$ upper limit in the remnant's mass, the power-law model does not predict any event for bosons with masses $m_s\lesssim 3\times 10^{-13}$. We obtain slightly smaller event rates than Ref.~\cite{Arvanitaki:2016qwi}, probably because we use a numerical calculation of the GW strain amplitude and more realistic assumptions on the BH mass and spin distributions (in~\cite{Arvanitaki:2016qwi} the progenitor BHs were assumed to have equal, aligned initial spins and equal masses, yielding a larger fraction of remnant BHs with spins $\chi>0.7$).

As pointed out in the previous section, the existence of a stochastic background from unresolved sources could produce a ``self-confusion noise'' and significantly reduce the rates, especially in the range of masses around $m_s\sim 3\times 10^{-13}$ (cf. the right panel of Fig.~\ref{fig:rates}). Here we have used a very optimistic scenario for the amplitude of the stochastic background, as described in the previous section. The real impact of the stochastic background will likely range somewhere between the two panels shown in Fig.~\ref{fig:rates}, depending on the astrophysical spin distribution of invidual BHs~\cite{Brito:2017wnc}.

As in~\cite{Arvanitaki:2016qwi,Baryakhtar:2017ngi}, we have assumed that superradiance does not operate during the early evolution of the binary. If superradiance is effective before merger and the cloud completely dissipates before merger, the binary members will have small spin and the remnant BH spin distribution will be highly peaked around $\chi\sim 0.68$~\cite{Buonanno:2007sv,Berti:2007fi}, so few merger remnants will have large enough spin to produce a detectable signal~\cite{Arvanitaki:2016qwi}. If one or both of the progenitor BHs are surrounded by a cloud, a full numerical evolution is necessary to determine the final state of the cloud(s) and its impact on the post-merger GW emission (but see~\cite{Baumann:2018vus} for analytical estimates).

\section{Final remarks}\label{sec:conclusions}

We have studied the parameter space that could be probed by GWs emitted by a cloud of ultralight bosons around a binary BH merger remnant at current and future ground-based GW detectors. Although most sources are expected to be too far away for these sources to be detectable by current ground-based GW detectors, we have shown that the prospects for future ground-based GW detectors are much more promising. The range of scalar field masses that can be probed overlaps with the range of masses that could be detected/constrained by all-sky searches for continuous GWs from isolated BHs~\cite{Arvanitaki:2014wva,Brito:2017zvb,Brito:2017wnc} and stochastic background searches~\cite{Brito:2017zvb,Brito:2017wnc}. However, constraints from a follow-up search would be independent of the assumptions made on the BH population, and therefore very robust against astrophysical uncertainties. In addition, a detection from a follow-up search would be a conclusive confirmation that the signal is emitted by a superradiant source, and therefore such a search would be complementary to observations from other channels. We have shown that, in the most optimistic scenario, we may expect Cosmic Explorer to detect dozens of binary BH mergers that would be ideal candidates to either detect or constrain the existence of  ultralight scalar fields.

We only considered scalar fields. Non-relativistic approximations~\cite{Baryakhtar:2017ngi} and numerical relativity simulations~\cite{East:2017mrj,East:2018glu} suggest that for vector fields the prospects to detect such signals could be significantly better.\footnote{We note that massive tensor fields are also prone to superradiant instabilities~\cite{Brito:2013wya}. However, in this case the non-linear evolution of the superradiant instability and subsequent GW emission should be described within a nonlinear theory of massive gravity.} However, since the formalism used in this work to compute the GW emission has not yet been implemented for vector fields, and is considerably more difficult to handle, we leave a more detailed analysis of vector fields for future work.

\begin{figure*}[t]
\begin{tabular}{cc}
\includegraphics[scale=0.386]{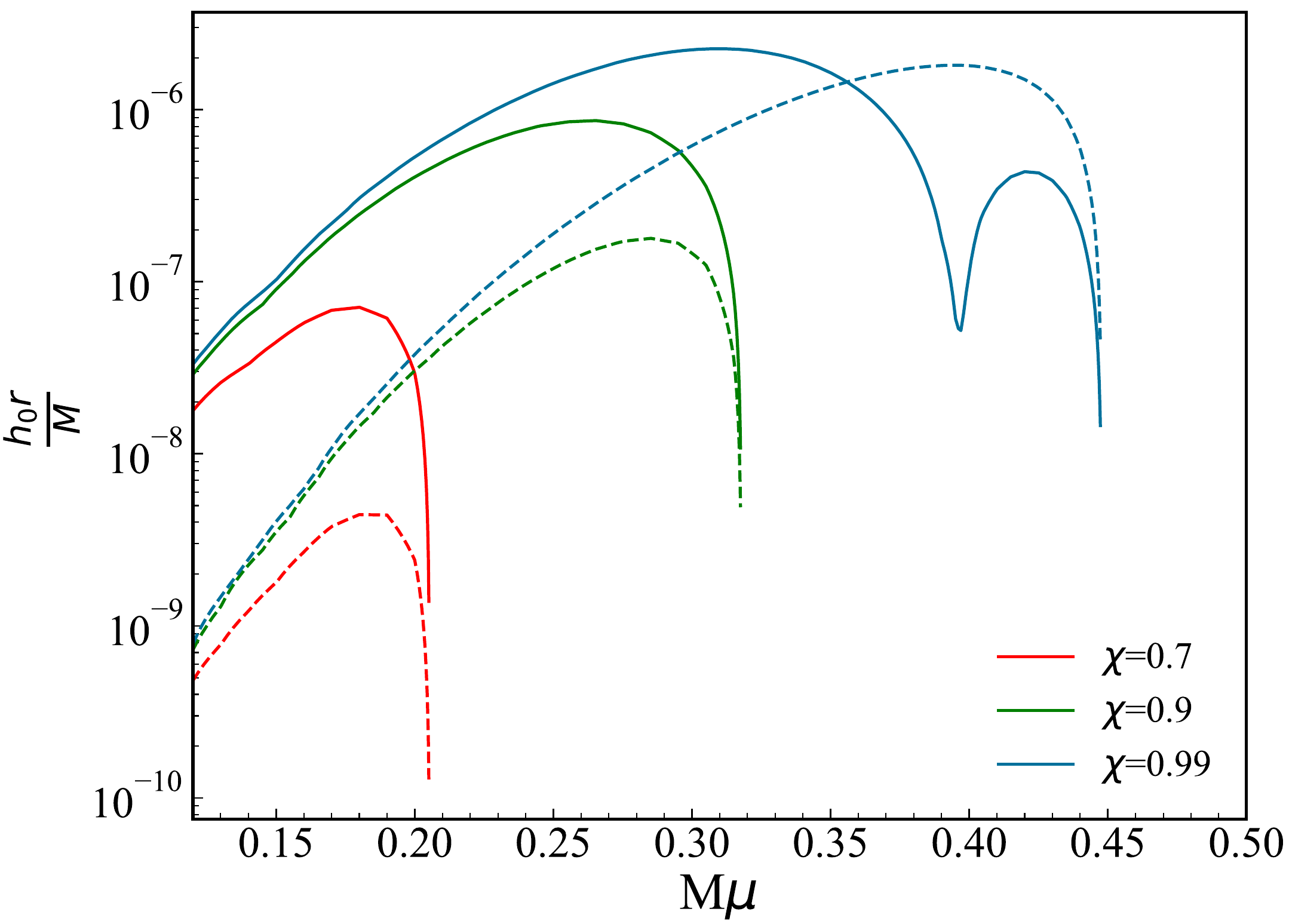}
\includegraphics[scale=0.386]{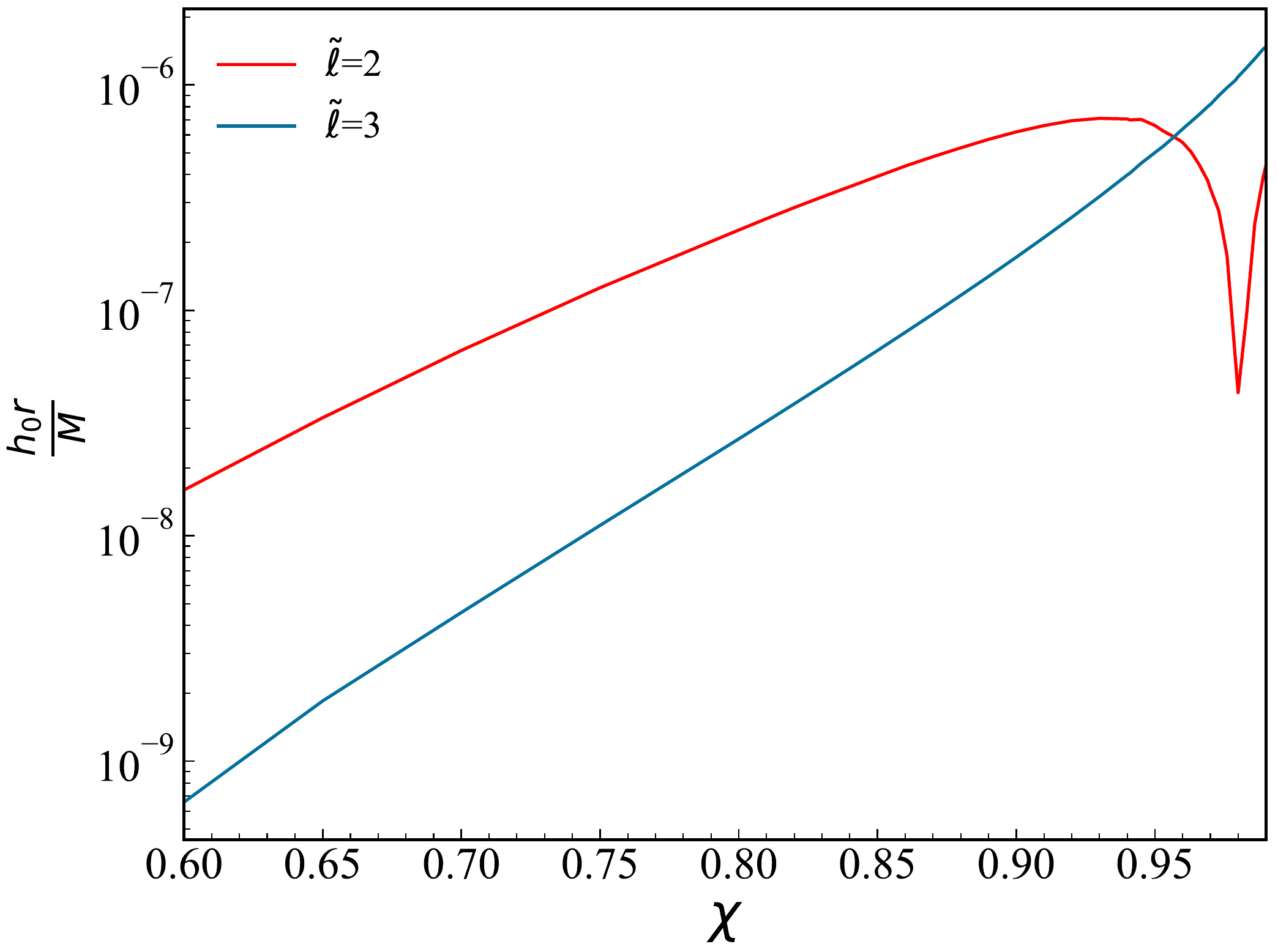}
\end{tabular}
\label{fig:l2l3}
\caption{Left panel: GW intrinsic amplitude $h_0^{(\tilde \ell)}r/M$ [c.f. eq.~\eqref{amplitude}] for the $\tilde \ell=2$ (solid lines) and $\tilde \ell=3$ (dashed lines) modes as a function of $M\mu$ and different selected values of the BH spin ($\chi=0.7\,, 0.9\,, 0.99$). Right panel: GW intrinsic amplitude for the $\tilde \ell=2$ and $\tilde \ell=3$ modes computed at the value of $M\mu$ that maximizes the superradiant instability growth rate (c.f Fig.~\ref{peakmu}).}
\end{figure*} 

\vspace{-0.1cm}
\acknowledgments
We thank Max Isi and Ling Sun for useful discussions. S.G. and E.B. are supported by NSF Grants No. PHY-1841464 and AST-1841358, and by NASA ATP Grant No. 17-ATP17-0225.
R.B. acknowledges financial support from the European Union's Horizon 2020 research and innovation programme under the Marie Sk\l odowska-Curie grant agreement No. 792862.
M.R. acknowledges support from the S\~ao Paulo Research Foundation (FAPESP, Brazil),
Grant No. 2013/09357-9, from Conselho Nacional de Desenvolvimento
Cient\'ifico e Tecnol\'ogico (CNPq, Brazil), Grant No. FA 309749/2017-4, and from the Fulbright Visiting Scholars Program.
 This work has received funding from the European Union’s Horizon 2020 research and innovation programme under the Marie Skłodowska-Curie grant agreement No. 690904.

\appendix
\section{Effect of higher multipoles on the radiation}\label{app:modes}

The energy levels of the BH/cloud system resemble the familiar structure of the hydrogen atom in quantum mechanics. Our estimates assume that only the $\ell=m=1$ mode of the scalar field gets populated by superradiant instabilities. For this mode, the nonaxisymmetric cloud emits gravitational radiation in all multipolar components with $\tilde{\ell} \ge 2\ell = 2$, but fixed $\tilde{m}=2m=2$~\cite{Yoshino:2013ofa}. Here we have only included the contribution of the $\tilde \ell=2$ mode in our calculations.

Here we show that, for the most important region of the parameter space, the mode with $\tilde \ell = 3$ and higher-order modes can indeed be neglected. In the left panel of Fig.~\ref{fig:l2l3} we show the GW intrinsic amplitude for the $\tilde \ell=2$ and $\tilde \ell=3$ modes as a function of $M\mu$ for selected values of the BH spin ($\chi=0.7\,, 0.9\,, 0.99$). In the right panel we show the contribution of each mode computed at the value of $M\mu$ that maximizes the superradiant instability growth rate as a function of $\chi$ (cf. Fig.~\ref{peakmu}). We do not show modes with $\tilde \ell>3$ because they have been shown to be subdominant relative to the $\tilde l=2$ and $\tilde l=3$ modes for any value of the BH spin and $M\mu$~\cite{Yoshino:2013ofa}. The emission is dominated by the $\tilde \ell=2$ mode up to $M\mu\simeq 0.35$ and spins $\sim 0.95$. On the other hand $\tilde l=3$ becomes important, and in fact dominates the emission, for spins $\chi\gtrsim 0.95$ and $M\mu\gtrsim 0.35$. However, it is extremely unlikely for BH mergers to produce remnants with spins higher than 0.95 unless the binary mass ratio is extremely large~\cite{Berti:2008af}, therefore our restriction to the $\tilde \ell=2$ mode is justified.

We remark in closing that our GW emissions estimates are conservative, because including high-order modes would yield higher GW amplitudes, at the cost of complicating both the theoretical analysis and signal searches.

\bibliographystyle{apsrev4-1}
\bibliography{refer}

\begin{thebibliography}{76}%
\makeatletter
\providecommand \@ifxundefined [1]{%
 \@ifx{#1\undefined}
}%
\providecommand \@ifnum [1]{%
 \ifnum #1\expandafter \@firstoftwo
 \else \expandafter \@secondoftwo
 \fi
}%
\providecommand \@ifx [1]{%
 \ifx #1\expandafter \@firstoftwo
 \else \expandafter \@secondoftwo
 \fi
}%
\providecommand \natexlab [1]{#1}%
\providecommand \enquote  [1]{``#1''}%
\providecommand \bibnamefont  [1]{#1}%
\providecommand \bibfnamefont [1]{#1}%
\providecommand \citenamefont [1]{#1}%
\providecommand \href@noop [0]{\@secondoftwo}%
\providecommand \href [0]{\begingroup \@sanitize@url \@href}%
\providecommand \@href[1]{\@@startlink{#1}\@@href}%
\providecommand \@@href[1]{\endgroup#1\@@endlink}%
\providecommand \@sanitize@url [0]{\catcode `\\12\catcode `\$12\catcode
  `\&12\catcode `\#12\catcode `\^12\catcode `\_12\catcode `\%12\relax}%
\providecommand \@@startlink[1]{}%
\providecommand \@@endlink[0]{}%
\providecommand \url  [0]{\begingroup\@sanitize@url \@url }%
\providecommand \@url [1]{\endgroup\@href {#1}{\urlprefix }}%
\providecommand \urlprefix  [0]{URL }%
\providecommand \Eprint [0]{\href }%
\providecommand \doibase [0]{http://dx.doi.org/}%
\providecommand \selectlanguage [0]{\@gobble}%
\providecommand \bibinfo  [0]{\@secondoftwo}%
\providecommand \bibfield  [0]{\@secondoftwo}%
\providecommand \translation [1]{[#1]}%
\providecommand \BibitemOpen [0]{}%
\providecommand \bibitemStop [0]{}%
\providecommand \bibitemNoStop [0]{.\EOS\space}%
\providecommand \EOS [0]{\spacefactor3000\relax}%
\providecommand \BibitemShut  [1]{\csname bibitem#1\endcsname}%
\let\auto@bib@innerbib\@empty
\bibitem [{\citenamefont {Wilczek}(1978)}]{Wilczek:1977pj}%
  \BibitemOpen
  \bibfield  {author} {\bibinfo {author} {\bibfnamefont {F.}~\bibnamefont
  {Wilczek}},\ }\href {\doibase 10.1103/PhysRevLett.40.279} {\bibfield
  {journal} {\bibinfo  {journal} {Phys. Rev. Lett.}\ }\textbf {\bibinfo
  {volume} {40}},\ \bibinfo {pages} {279} (\bibinfo {year} {1978})}\BibitemShut
  {NoStop}%
\bibitem [{\citenamefont {Arvanitaki}\ \emph {et~al.}(2010)\citenamefont
  {Arvanitaki}, \citenamefont {Dimopoulos}, \citenamefont {Dubovsky},
  \citenamefont {Kaloper},\ and\ \citenamefont
  {March-Russell}}]{Arvanitaki:2009fg}%
  \BibitemOpen
  \bibfield  {author} {\bibinfo {author} {\bibfnamefont {A.}~\bibnamefont
  {Arvanitaki}}, \bibinfo {author} {\bibfnamefont {S.}~\bibnamefont
  {Dimopoulos}}, \bibinfo {author} {\bibfnamefont {S.}~\bibnamefont
  {Dubovsky}}, \bibinfo {author} {\bibfnamefont {N.}~\bibnamefont {Kaloper}}, \
  and\ \bibinfo {author} {\bibfnamefont {J.}~\bibnamefont {March-Russell}},\
  }\href {\doibase 10.1103/PhysRevD.81.123530} {\bibfield  {journal} {\bibinfo
  {journal} {Phys. Rev.}\ }\textbf {\bibinfo {volume} {D81}},\ \bibinfo {pages}
  {123530} (\bibinfo {year} {2010})},\ \Eprint {http://arxiv.org/abs/0905.4720}
  {arXiv:0905.4720 [hep-th]} \BibitemShut {NoStop}%
\bibitem [{\citenamefont {Bertone}\ \emph {et~al.}(2005)\citenamefont
  {Bertone}, \citenamefont {Hooper},\ and\ \citenamefont
  {Silk}}]{Bertone:2004pz}%
  \BibitemOpen
  \bibfield  {author} {\bibinfo {author} {\bibfnamefont {G.}~\bibnamefont
  {Bertone}}, \bibinfo {author} {\bibfnamefont {D.}~\bibnamefont {Hooper}}, \
  and\ \bibinfo {author} {\bibfnamefont {J.}~\bibnamefont {Silk}},\ }\href
  {\doibase 10.1016/j.physrep.2004.08.031} {\bibfield  {journal} {\bibinfo
  {journal} {Phys. Rept.}\ }\textbf {\bibinfo {volume} {405}},\ \bibinfo
  {pages} {279} (\bibinfo {year} {2005})},\ \Eprint
  {http://arxiv.org/abs/hep-ph/0404175} {arXiv:hep-ph/0404175 [hep-ph]}
  \BibitemShut {NoStop}%
\bibitem [{\citenamefont {Marsh}(2016)}]{Marsh:2015xka}%
  \BibitemOpen
  \bibfield  {author} {\bibinfo {author} {\bibfnamefont {D.~J.~E.}\
  \bibnamefont {Marsh}},\ }\href {\doibase 10.1016/j.physrep.2016.06.005}
  {\bibfield  {journal} {\bibinfo  {journal} {Phys. Rept.}\ }\textbf {\bibinfo
  {volume} {643}},\ \bibinfo {pages} {1} (\bibinfo {year} {2016})},\ \Eprint
  {http://arxiv.org/abs/1510.07633} {arXiv:1510.07633 [astro-ph.CO]}
  \BibitemShut {NoStop}%
\bibitem [{\citenamefont {Poulin}\ \emph {et~al.}(2018)\citenamefont {Poulin},
  \citenamefont {Smith}, \citenamefont {Grin}, \citenamefont {Karwal},\ and\
  \citenamefont {Kamionkowski}}]{Poulin:2018dzj}%
  \BibitemOpen
  \bibfield  {author} {\bibinfo {author} {\bibfnamefont {V.}~\bibnamefont
  {Poulin}}, \bibinfo {author} {\bibfnamefont {T.~L.}\ \bibnamefont {Smith}},
  \bibinfo {author} {\bibfnamefont {D.}~\bibnamefont {Grin}}, \bibinfo {author}
  {\bibfnamefont {T.}~\bibnamefont {Karwal}}, \ and\ \bibinfo {author}
  {\bibfnamefont {M.}~\bibnamefont {Kamionkowski}},\ }\href@noop {} {\
  (\bibinfo {year} {2018})},\ \Eprint {http://arxiv.org/abs/1806.10608}
  {arXiv:1806.10608 [astro-ph.CO]} \BibitemShut {NoStop}%
\bibitem [{\citenamefont {Arvanitaki}\ and\ \citenamefont
  {Dubovsky}(2011)}]{Arvanitaki:2010sy}%
  \BibitemOpen
  \bibfield  {author} {\bibinfo {author} {\bibfnamefont {A.}~\bibnamefont
  {Arvanitaki}}\ and\ \bibinfo {author} {\bibfnamefont {S.}~\bibnamefont
  {Dubovsky}},\ }\href {\doibase 10.1103/PhysRevD.83.044026} {\bibfield
  {journal} {\bibinfo  {journal} {Phys. Rev.}\ }\textbf {\bibinfo {volume}
  {D83}},\ \bibinfo {pages} {044026} (\bibinfo {year} {2011})},\ \Eprint
  {http://arxiv.org/abs/1004.3558} {arXiv:1004.3558 [hep-th]} \BibitemShut
  {NoStop}%
\bibitem [{\citenamefont {Arvanitaki}\ \emph {et~al.}(2015)\citenamefont
  {Arvanitaki}, \citenamefont {Baryakhtar},\ and\ \citenamefont
  {Huang}}]{Arvanitaki:2014wva}%
  \BibitemOpen
  \bibfield  {author} {\bibinfo {author} {\bibfnamefont {A.}~\bibnamefont
  {Arvanitaki}}, \bibinfo {author} {\bibfnamefont {M.}~\bibnamefont
  {Baryakhtar}}, \ and\ \bibinfo {author} {\bibfnamefont {X.}~\bibnamefont
  {Huang}},\ }\href {\doibase 10.1103/PhysRevD.91.084011} {\bibfield  {journal}
  {\bibinfo  {journal} {Phys. Rev.}\ }\textbf {\bibinfo {volume} {D91}},\
  \bibinfo {pages} {084011} (\bibinfo {year} {2015})},\ \Eprint
  {http://arxiv.org/abs/1411.2263} {arXiv:1411.2263 [hep-ph]} \BibitemShut
  {NoStop}%
\bibitem [{\citenamefont {Brito}\ \emph
  {et~al.}(2015{\natexlab{a}})\citenamefont {Brito}, \citenamefont {Cardoso},\
  and\ \citenamefont {Pani}}]{Brito:2014wla}%
  \BibitemOpen
  \bibfield  {author} {\bibinfo {author} {\bibfnamefont {R.}~\bibnamefont
  {Brito}}, \bibinfo {author} {\bibfnamefont {V.}~\bibnamefont {Cardoso}}, \
  and\ \bibinfo {author} {\bibfnamefont {P.}~\bibnamefont {Pani}},\ }\href
  {\doibase 10.1088/0264-9381/32/13/134001} {\bibfield  {journal} {\bibinfo
  {journal} {Class. Quant. Grav.}\ }\textbf {\bibinfo {volume} {32}},\ \bibinfo
  {pages} {134001} (\bibinfo {year} {2015}{\natexlab{a}})},\ \Eprint
  {http://arxiv.org/abs/1411.0686} {arXiv:1411.0686 [gr-qc]} \BibitemShut
  {NoStop}%
\bibitem [{\citenamefont {Arvanitaki}\ \emph {et~al.}(2017)\citenamefont
  {Arvanitaki}, \citenamefont {Baryakhtar}, \citenamefont {Dimopoulos},
  \citenamefont {Dubovsky},\ and\ \citenamefont
  {Lasenby}}]{Arvanitaki:2016qwi}%
  \BibitemOpen
  \bibfield  {author} {\bibinfo {author} {\bibfnamefont {A.}~\bibnamefont
  {Arvanitaki}}, \bibinfo {author} {\bibfnamefont {M.}~\bibnamefont
  {Baryakhtar}}, \bibinfo {author} {\bibfnamefont {S.}~\bibnamefont
  {Dimopoulos}}, \bibinfo {author} {\bibfnamefont {S.}~\bibnamefont
  {Dubovsky}}, \ and\ \bibinfo {author} {\bibfnamefont {R.}~\bibnamefont
  {Lasenby}},\ }\href {\doibase 10.1103/PhysRevD.95.043001} {\bibfield
  {journal} {\bibinfo  {journal} {Phys. Rev.}\ }\textbf {\bibinfo {volume}
  {D95}},\ \bibinfo {pages} {043001} (\bibinfo {year} {2017})},\ \Eprint
  {http://arxiv.org/abs/1604.03958} {arXiv:1604.03958 [hep-ph]} \BibitemShut
  {NoStop}%
\bibitem [{\citenamefont {Baryakhtar}\ \emph {et~al.}(2017)\citenamefont
  {Baryakhtar}, \citenamefont {Lasenby},\ and\ \citenamefont
  {Teo}}]{Baryakhtar:2017ngi}%
  \BibitemOpen
  \bibfield  {author} {\bibinfo {author} {\bibfnamefont {M.}~\bibnamefont
  {Baryakhtar}}, \bibinfo {author} {\bibfnamefont {R.}~\bibnamefont {Lasenby}},
  \ and\ \bibinfo {author} {\bibfnamefont {M.}~\bibnamefont {Teo}},\ }\href
  {\doibase 10.1103/PhysRevD.96.035019} {\bibfield  {journal} {\bibinfo
  {journal} {Phys. Rev.}\ }\textbf {\bibinfo {volume} {D96}},\ \bibinfo {pages}
  {035019} (\bibinfo {year} {2017})},\ \Eprint
  {http://arxiv.org/abs/1704.05081} {arXiv:1704.05081 [hep-ph]} \BibitemShut
  {NoStop}%
\bibitem [{\citenamefont {Brito}\ \emph
  {et~al.}(2017{\natexlab{a}})\citenamefont {Brito}, \citenamefont {Ghosh},
  \citenamefont {Barausse}, \citenamefont {Berti}, \citenamefont {Cardoso},
  \citenamefont {Dvorkin}, \citenamefont {Klein},\ and\ \citenamefont
  {Pani}}]{Brito:2017wnc}%
  \BibitemOpen
  \bibfield  {author} {\bibinfo {author} {\bibfnamefont {R.}~\bibnamefont
  {Brito}}, \bibinfo {author} {\bibfnamefont {S.}~\bibnamefont {Ghosh}},
  \bibinfo {author} {\bibfnamefont {E.}~\bibnamefont {Barausse}}, \bibinfo
  {author} {\bibfnamefont {E.}~\bibnamefont {Berti}}, \bibinfo {author}
  {\bibfnamefont {V.}~\bibnamefont {Cardoso}}, \bibinfo {author} {\bibfnamefont
  {I.}~\bibnamefont {Dvorkin}}, \bibinfo {author} {\bibfnamefont
  {A.}~\bibnamefont {Klein}}, \ and\ \bibinfo {author} {\bibfnamefont
  {P.}~\bibnamefont {Pani}},\ }\href {\doibase 10.1103/PhysRevLett.119.131101}
  {\bibfield  {journal} {\bibinfo  {journal} {Phys. Rev. Lett.}\ }\textbf
  {\bibinfo {volume} {119}},\ \bibinfo {pages} {131101} (\bibinfo {year}
  {2017}{\natexlab{a}})},\ \Eprint {http://arxiv.org/abs/1706.05097}
  {arXiv:1706.05097 [gr-qc]} \BibitemShut {NoStop}%
\bibitem [{\citenamefont {Brito}\ \emph
  {et~al.}(2017{\natexlab{b}})\citenamefont {Brito}, \citenamefont {Ghosh},
  \citenamefont {Barausse}, \citenamefont {Berti}, \citenamefont {Cardoso},
  \citenamefont {Dvorkin}, \citenamefont {Klein},\ and\ \citenamefont
  {Pani}}]{Brito:2017zvb}%
  \BibitemOpen
  \bibfield  {author} {\bibinfo {author} {\bibfnamefont {R.}~\bibnamefont
  {Brito}}, \bibinfo {author} {\bibfnamefont {S.}~\bibnamefont {Ghosh}},
  \bibinfo {author} {\bibfnamefont {E.}~\bibnamefont {Barausse}}, \bibinfo
  {author} {\bibfnamefont {E.}~\bibnamefont {Berti}}, \bibinfo {author}
  {\bibfnamefont {V.}~\bibnamefont {Cardoso}}, \bibinfo {author} {\bibfnamefont
  {I.}~\bibnamefont {Dvorkin}}, \bibinfo {author} {\bibfnamefont
  {A.}~\bibnamefont {Klein}}, \ and\ \bibinfo {author} {\bibfnamefont
  {P.}~\bibnamefont {Pani}},\ }\href {\doibase 10.1103/PhysRevD.96.064050}
  {\bibfield  {journal} {\bibinfo  {journal} {Phys. Rev.}\ }\textbf {\bibinfo
  {volume} {D96}},\ \bibinfo {pages} {064050} (\bibinfo {year}
  {2017}{\natexlab{b}})},\ \Eprint {http://arxiv.org/abs/1706.06311}
  {arXiv:1706.06311 [gr-qc]} \BibitemShut {NoStop}%
\bibitem [{\citenamefont {Baumann}\ \emph {et~al.}(2018)\citenamefont
  {Baumann}, \citenamefont {Chia},\ and\ \citenamefont
  {Porto}}]{Baumann:2018vus}%
  \BibitemOpen
  \bibfield  {author} {\bibinfo {author} {\bibfnamefont {D.}~\bibnamefont
  {Baumann}}, \bibinfo {author} {\bibfnamefont {H.~S.}\ \bibnamefont {Chia}}, \
  and\ \bibinfo {author} {\bibfnamefont {R.~A.}\ \bibnamefont {Porto}},\
  }\href@noop {} {\  (\bibinfo {year} {2018})},\ \Eprint
  {http://arxiv.org/abs/1804.03208} {arXiv:1804.03208 [gr-qc]} \BibitemShut
  {NoStop}%
\bibitem [{\citenamefont {Hannuksela}\ \emph {et~al.}(2018)\citenamefont
  {Hannuksela}, \citenamefont {Brito}, \citenamefont {Berti},\ and\
  \citenamefont {Li}}]{Hannuksela:2018izj}%
  \BibitemOpen
  \bibfield  {author} {\bibinfo {author} {\bibfnamefont {O.~A.}\ \bibnamefont
  {Hannuksela}}, \bibinfo {author} {\bibfnamefont {R.}~\bibnamefont {Brito}},
  \bibinfo {author} {\bibfnamefont {E.}~\bibnamefont {Berti}}, \ and\ \bibinfo
  {author} {\bibfnamefont {T.~G.~F.}\ \bibnamefont {Li}},\ }\href@noop {} {\
  (\bibinfo {year} {2018})},\ \Eprint {http://arxiv.org/abs/1804.09659}
  {arXiv:1804.09659 [astro-ph.HE]} \BibitemShut {NoStop}%
\bibitem [{\citenamefont {Isi}\ \emph {et~al.}(2018)\citenamefont {Isi},
  \citenamefont {Sun}, \citenamefont {Brito},\ and\ \citenamefont
  {Melatos}}]{Isi:2018pzk}%
  \BibitemOpen
  \bibfield  {author} {\bibinfo {author} {\bibfnamefont {M.}~\bibnamefont
  {Isi}}, \bibinfo {author} {\bibfnamefont {L.}~\bibnamefont {Sun}}, \bibinfo
  {author} {\bibfnamefont {R.}~\bibnamefont {Brito}}, \ and\ \bibinfo {author}
  {\bibfnamefont {A.}~\bibnamefont {Melatos}},\ }\href@noop {} {\  (\bibinfo
  {year} {2018})},\ \Eprint {http://arxiv.org/abs/1810.03812} {arXiv:1810.03812
  [gr-qc]} \BibitemShut {NoStop}%
\bibitem [{\citenamefont {Brito}\ \emph
  {et~al.}(2015{\natexlab{b}})\citenamefont {Brito}, \citenamefont {Cardoso},\
  and\ \citenamefont {Pani}}]{Brito:2015oca}%
  \BibitemOpen
  \bibfield  {author} {\bibinfo {author} {\bibfnamefont {R.}~\bibnamefont
  {Brito}}, \bibinfo {author} {\bibfnamefont {V.}~\bibnamefont {Cardoso}}, \
  and\ \bibinfo {author} {\bibfnamefont {P.}~\bibnamefont {Pani}},\ }\href
  {\doibase 10.1007/978-3-319-19000-6} {\bibfield  {journal} {\bibinfo
  {journal} {Lect. Notes Phys.}\ }\textbf {\bibinfo {volume} {906}},\ \bibinfo
  {pages} {pp.1} (\bibinfo {year} {2015}{\natexlab{b}})},\ \Eprint
  {http://arxiv.org/abs/1501.06570} {arXiv:1501.06570 [gr-qc]} \BibitemShut
  {NoStop}%
\bibitem [{\citenamefont {Detweiler}(1980)}]{Detweiler:1980uk}%
  \BibitemOpen
  \bibfield  {author} {\bibinfo {author} {\bibfnamefont {S.~L.}\ \bibnamefont
  {Detweiler}},\ }\href {\doibase 10.1103/PhysRevD.22.2323} {\bibfield
  {journal} {\bibinfo  {journal} {Phys. Rev.}\ }\textbf {\bibinfo {volume}
  {D22}},\ \bibinfo {pages} {2323} (\bibinfo {year} {1980})}\BibitemShut
  {NoStop}%
\bibitem [{\citenamefont {Dolan}(2007)}]{Dolan:2007mj}%
  \BibitemOpen
  \bibfield  {author} {\bibinfo {author} {\bibfnamefont {S.~R.}\ \bibnamefont
  {Dolan}},\ }\href {\doibase 10.1103/PhysRevD.76.084001} {\bibfield  {journal}
  {\bibinfo  {journal} {Phys. Rev.}\ }\textbf {\bibinfo {volume} {D76}},\
  \bibinfo {pages} {084001} (\bibinfo {year} {2007})},\ \Eprint
  {http://arxiv.org/abs/0705.2880} {arXiv:0705.2880 [gr-qc]} \BibitemShut
  {NoStop}%
\bibitem [{\citenamefont {Christodoulou}\ and\ \citenamefont
  {Ruffini}(1971)}]{Christodoulou:1972kt}%
  \BibitemOpen
  \bibfield  {author} {\bibinfo {author} {\bibfnamefont {D.}~\bibnamefont
  {Christodoulou}}\ and\ \bibinfo {author} {\bibfnamefont {R.}~\bibnamefont
  {Ruffini}},\ }\href {\doibase 10.1103/PhysRevD.4.3552} {\bibfield  {journal}
  {\bibinfo  {journal} {Phys. Rev.}\ }\textbf {\bibinfo {volume} {D4}},\
  \bibinfo {pages} {3552} (\bibinfo {year} {1971})}\BibitemShut {NoStop}%
\bibitem [{\citenamefont {Herdeiro}\ and\ \citenamefont
  {Radu}(2017)}]{Herdeiro:2017phl}%
  \BibitemOpen
  \bibfield  {author} {\bibinfo {author} {\bibfnamefont {C.~A.~R.}\
  \bibnamefont {Herdeiro}}\ and\ \bibinfo {author} {\bibfnamefont
  {E.}~\bibnamefont {Radu}},\ }\href {\doibase 10.1103/PhysRevLett.119.261101}
  {\bibfield  {journal} {\bibinfo  {journal} {Phys. Rev. Lett.}\ }\textbf
  {\bibinfo {volume} {119}},\ \bibinfo {pages} {261101} (\bibinfo {year}
  {2017})},\ \Eprint {http://arxiv.org/abs/1706.06597} {arXiv:1706.06597
  [gr-qc]} \BibitemShut {NoStop}%
\bibitem [{\citenamefont {East}\ and\ \citenamefont
  {Pretorius}(2017)}]{East:2017ovw}%
  \BibitemOpen
  \bibfield  {author} {\bibinfo {author} {\bibfnamefont {W.~E.}\ \bibnamefont
  {East}}\ and\ \bibinfo {author} {\bibfnamefont {F.}~\bibnamefont
  {Pretorius}},\ }\href {\doibase 10.1103/PhysRevLett.119.041101} {\bibfield
  {journal} {\bibinfo  {journal} {Phys. Rev. Lett.}\ }\textbf {\bibinfo
  {volume} {119}},\ \bibinfo {pages} {041101} (\bibinfo {year} {2017})},\
  \Eprint {http://arxiv.org/abs/1704.04791} {arXiv:1704.04791 [gr-qc]}
  \BibitemShut {NoStop}%
\bibitem [{\citenamefont {East}(2018)}]{East:2018glu}%
  \BibitemOpen
  \bibfield  {author} {\bibinfo {author} {\bibfnamefont {W.~E.}\ \bibnamefont
  {East}},\ }\href {\doibase 10.1103/PhysRevLett.121.131104} {\bibfield
  {journal} {\bibinfo  {journal} {Phys. Rev. Lett.}\ }\textbf {\bibinfo
  {volume} {121}},\ \bibinfo {pages} {131104} (\bibinfo {year} {2018})},\
  \Eprint {http://arxiv.org/abs/1807.00043} {arXiv:1807.00043 [gr-qc]}
  \BibitemShut {NoStop}%
\bibitem [{\citenamefont {Yoshino}\ and\ \citenamefont
  {Kodama}(2012)}]{Yoshino:2012kn}%
  \BibitemOpen
  \bibfield  {author} {\bibinfo {author} {\bibfnamefont {H.}~\bibnamefont
  {Yoshino}}\ and\ \bibinfo {author} {\bibfnamefont {H.}~\bibnamefont
  {Kodama}},\ }\href {\doibase 10.1143/PTP.128.153} {\bibfield  {journal}
  {\bibinfo  {journal} {Prog. Theor. Phys.}\ }\textbf {\bibinfo {volume}
  {128}},\ \bibinfo {pages} {153} (\bibinfo {year} {2012})},\ \Eprint
  {http://arxiv.org/abs/1203.5070} {arXiv:1203.5070 [gr-qc]} \BibitemShut
  {NoStop}%
\bibitem [{\citenamefont {Rosa}\ and\ \citenamefont
  {Kephart}(2018)}]{Rosa:2017ury}%
  \BibitemOpen
  \bibfield  {author} {\bibinfo {author} {\bibfnamefont {J.~G.}\ \bibnamefont
  {Rosa}}\ and\ \bibinfo {author} {\bibfnamefont {T.~W.}\ \bibnamefont
  {Kephart}},\ }\href {\doibase 10.1103/PhysRevLett.120.231102} {\bibfield
  {journal} {\bibinfo  {journal} {Phys. Rev. Lett.}\ }\textbf {\bibinfo
  {volume} {120}},\ \bibinfo {pages} {231102} (\bibinfo {year} {2018})},\
  \Eprint {http://arxiv.org/abs/1709.06581} {arXiv:1709.06581 [gr-qc]}
  \BibitemShut {NoStop}%
\bibitem [{\citenamefont {Boskovic}\ \emph {et~al.}(2018)\citenamefont
  {Boskovic}, \citenamefont {Brito}, \citenamefont {Cardoso}, \citenamefont
  {Ikeda},\ and\ \citenamefont {Witek}}]{Boskovic:2018lkj}%
  \BibitemOpen
  \bibfield  {author} {\bibinfo {author} {\bibfnamefont {M.}~\bibnamefont
  {Boskovic}}, \bibinfo {author} {\bibfnamefont {R.}~\bibnamefont {Brito}},
  \bibinfo {author} {\bibfnamefont {V.}~\bibnamefont {Cardoso}}, \bibinfo
  {author} {\bibfnamefont {T.}~\bibnamefont {Ikeda}}, \ and\ \bibinfo {author}
  {\bibfnamefont {H.}~\bibnamefont {Witek}},\ }\href@noop {} {\  (\bibinfo
  {year} {2018})},\ \Eprint {http://arxiv.org/abs/1811.04945} {arXiv:1811.04945
  [gr-qc]} \BibitemShut {NoStop}%
\bibitem [{\citenamefont {Ikeda}\ \emph {et~al.}(2018)\citenamefont {Ikeda},
  \citenamefont {Brito},\ and\ \citenamefont {Cardoso}}]{Ikeda:2018nhb}%
  \BibitemOpen
  \bibfield  {author} {\bibinfo {author} {\bibfnamefont {T.}~\bibnamefont
  {Ikeda}}, \bibinfo {author} {\bibfnamefont {R.}~\bibnamefont {Brito}}, \ and\
  \bibinfo {author} {\bibfnamefont {V.}~\bibnamefont {Cardoso}},\ }\href@noop
  {} {\  (\bibinfo {year} {2018})},\ \Eprint {http://arxiv.org/abs/1811.04950}
  {arXiv:1811.04950 [gr-qc]} \BibitemShut {NoStop}%
\bibitem [{\citenamefont {Cardoso}\ \emph {et~al.}(2018)\citenamefont
  {Cardoso}, \citenamefont {Dias}, \citenamefont {Hartnett}, \citenamefont
  {Middleton}, \citenamefont {Pani},\ and\ \citenamefont
  {Santos}}]{Cardoso:2018tly}%
  \BibitemOpen
  \bibfield  {author} {\bibinfo {author} {\bibfnamefont {V.}~\bibnamefont
  {Cardoso}}, \bibinfo {author} {\bibfnamefont {O.~J.~C.}\ \bibnamefont
  {Dias}}, \bibinfo {author} {\bibfnamefont {G.~S.}\ \bibnamefont {Hartnett}},
  \bibinfo {author} {\bibfnamefont {M.}~\bibnamefont {Middleton}}, \bibinfo
  {author} {\bibfnamefont {P.}~\bibnamefont {Pani}}, \ and\ \bibinfo {author}
  {\bibfnamefont {J.~E.}\ \bibnamefont {Santos}},\ }\href {\doibase
  10.1088/1475-7516/2018/03/043} {\bibfield  {journal} {\bibinfo  {journal}
  {JCAP}\ }\textbf {\bibinfo {volume} {1803}},\ \bibinfo {pages} {043}
  (\bibinfo {year} {2018})},\ \Eprint {http://arxiv.org/abs/1801.01420}
  {arXiv:1801.01420 [gr-qc]} \BibitemShut {NoStop}%
\bibitem [{\citenamefont {Krawczynski}(2018)}]{Krawczynski:2018fnw}%
  \BibitemOpen
  \bibfield  {author} {\bibinfo {author} {\bibfnamefont {H.}~\bibnamefont
  {Krawczynski}},\ }\href {\doibase 10.1007/s10714-018-2419-8} {\bibfield
  {journal} {\bibinfo  {journal} {Gen. Rel. Grav.}\ }\textbf {\bibinfo {volume}
  {50}},\ \bibinfo {pages} {100} (\bibinfo {year} {2018})},\ \Eprint
  {http://arxiv.org/abs/1806.10347} {arXiv:1806.10347 [astro-ph.HE]}
  \BibitemShut {NoStop}%
\bibitem [{\citenamefont {{The LIGO Scientific Collaboration and The Virgo
  Collaboration}}(2018{\natexlab{a}})}]{LIGOScientific:2018mvr}%
  \BibitemOpen
  \bibfield  {author} {\bibinfo {author} {\bibnamefont {{The LIGO Scientific
  Collaboration and The Virgo Collaboration}}},\ }\href@noop {} {\  (\bibinfo
  {year} {2018}{\natexlab{a}})},\ \Eprint {http://arxiv.org/abs/1811.12907}
  {arXiv:1811.12907 [astro-ph.HE]} \BibitemShut {NoStop}%
\bibitem [{\citenamefont {{LIGO Scientific
  Collaboration}}(2018{\natexlab{a}})}]{LIGOcurve}%
  \BibitemOpen
  \bibfield  {author} {\bibinfo {author} {\bibnamefont {{LIGO Scientific
  Collaboration}}},\ }\href@noop {} {\bibfield  {journal} {\bibinfo  {journal}
  {\href{https://dcc.ligo.org/T1800044/public}{dcc.ligo.org/T1800044}}\ }
  (\bibinfo {year} {2018}{\natexlab{a}})}\BibitemShut {NoStop}%
\bibitem [{\citenamefont {{LIGO Scientific
  Collaboration}}(2018{\natexlab{b}})}]{Apluscurve}%
  \BibitemOpen
  \bibfield  {author} {\bibinfo {author} {\bibnamefont {{LIGO Scientific
  Collaboration}}},\ }\href@noop {} {\bibfield  {journal} {\bibinfo  {journal}
  {\href{https://dcc.ligo.org/LIGO-T1800042/public}{dcc.ligo.org/LIGO-T1800042}}\
  } (\bibinfo {year} {2018}{\natexlab{b}})}\BibitemShut {NoStop}%
\bibitem [{\citenamefont {{LIGO Scientific
  Collaboration}}(2016)}]{Voyagercurve}%
  \BibitemOpen
  \bibfield  {author} {\bibinfo {author} {\bibnamefont {{LIGO Scientific
  Collaboration}}},\ }\href@noop {} {\bibfield  {journal} {\bibinfo  {journal}
  {\href{https://dcc.ligo.org/LIGO-T1500290/public}{dcc.ligo.org/LIGO-T1500290}}\
  } (\bibinfo {year} {2016})}\BibitemShut {NoStop}%
\bibitem [{\citenamefont {{Abbott}}\ \emph {et~al.}(2017)\citenamefont
  {{Abbott}}, \citenamefont {{Abbott}}, \citenamefont {{Abbott}}, \citenamefont
  {{Abernathy}}, \citenamefont {{Ackley}}, \citenamefont {{Adams}},
  \citenamefont {{Addesso}}, \citenamefont {{Adhikari}}, \citenamefont
  {{Adya}}, \citenamefont {{Affeldt}},\ and\ \citenamefont
  {et~al.}}]{2017CQGra..34d4001A}%
  \BibitemOpen
  \bibfield  {author} {\bibinfo {author} {\bibfnamefont {B.~P.}\ \bibnamefont
  {{Abbott}}}, \bibinfo {author} {\bibfnamefont {R.}~\bibnamefont {{Abbott}}},
  \bibinfo {author} {\bibfnamefont {T.~D.}\ \bibnamefont {{Abbott}}}, \bibinfo
  {author} {\bibfnamefont {M.~R.}\ \bibnamefont {{Abernathy}}}, \bibinfo
  {author} {\bibfnamefont {K.}~\bibnamefont {{Ackley}}}, \bibinfo {author}
  {\bibfnamefont {C.}~\bibnamefont {{Adams}}}, \bibinfo {author} {\bibfnamefont
  {P.}~\bibnamefont {{Addesso}}}, \bibinfo {author} {\bibfnamefont {R.~X.}\
  \bibnamefont {{Adhikari}}}, \bibinfo {author} {\bibfnamefont {V.~B.}\
  \bibnamefont {{Adya}}}, \bibinfo {author} {\bibfnamefont {C.}~\bibnamefont
  {{Affeldt}}}, \ and\ \bibinfo {author} {\bibnamefont {et~al.}},\ }\href
  {\doibase 10.1088/1361-6382/aa51f4} {\bibfield  {journal} {\bibinfo
  {journal} {Classical and Quantum Gravity}\ }\textbf {\bibinfo {volume}
  {34}},\ \bibinfo {eid} {044001} (\bibinfo {year} {2017})},\ \Eprint
  {http://arxiv.org/abs/1607.08697} {arXiv:1607.08697 [astro-ph.IM]}
  \BibitemShut {NoStop}%
\bibitem [{\citenamefont {Berti}\ \emph {et~al.}(2006)\citenamefont {Berti},
  \citenamefont {Cardoso},\ and\ \citenamefont {Casals}}]{Berti:2005gp}%
  \BibitemOpen
  \bibfield  {author} {\bibinfo {author} {\bibfnamefont {E.}~\bibnamefont
  {Berti}}, \bibinfo {author} {\bibfnamefont {V.}~\bibnamefont {Cardoso}}, \
  and\ \bibinfo {author} {\bibfnamefont {M.}~\bibnamefont {Casals}},\ }\href
  {\doibase 10.1103/PhysRevD.73.109902, 10.1103/PhysRevD.73.024013} {\bibfield
  {journal} {\bibinfo  {journal} {Phys. Rev.}\ }\textbf {\bibinfo {volume}
  {D73}},\ \bibinfo {pages} {024013} (\bibinfo {year} {2006})},\ \bibinfo
  {note} {[Erratum: Phys. Rev.D73,109902(2006)]},\ \Eprint
  {http://arxiv.org/abs/gr-qc/0511111} {arXiv:gr-qc/0511111 [gr-qc]}
  \BibitemShut {NoStop}%
\bibitem [{\citenamefont {Brill}\ \emph {et~al.}(1972)\citenamefont {Brill},
  \citenamefont {Chrzanowski}, \citenamefont {Martin~Pereira}, \citenamefont
  {Fackerell},\ and\ \citenamefont {Ipser}}]{Brill:1972xj}%
  \BibitemOpen
  \bibfield  {author} {\bibinfo {author} {\bibfnamefont {D.~R.}\ \bibnamefont
  {Brill}}, \bibinfo {author} {\bibfnamefont {P.~L.}\ \bibnamefont
  {Chrzanowski}}, \bibinfo {author} {\bibfnamefont {C.}~\bibnamefont
  {Martin~Pereira}}, \bibinfo {author} {\bibfnamefont {E.~D.}\ \bibnamefont
  {Fackerell}}, \ and\ \bibinfo {author} {\bibfnamefont {J.~R.}\ \bibnamefont
  {Ipser}},\ }\href {\doibase 10.1103/PhysRevD.5.1913} {\bibfield  {journal}
  {\bibinfo  {journal} {Phys. Rev.}\ }\textbf {\bibinfo {volume} {D5}},\
  \bibinfo {pages} {1913} (\bibinfo {year} {1972})}\BibitemShut {NoStop}%
\bibitem [{\citenamefont {Cardoso}\ and\ \citenamefont
  {Yoshida}(2005)}]{Cardoso:2005vk}%
  \BibitemOpen
  \bibfield  {author} {\bibinfo {author} {\bibfnamefont {V.}~\bibnamefont
  {Cardoso}}\ and\ \bibinfo {author} {\bibfnamefont {S.}~\bibnamefont
  {Yoshida}},\ }\href {\doibase 10.1088/1126-6708/2005/07/009} {\bibfield
  {journal} {\bibinfo  {journal} {JHEP}\ }\textbf {\bibinfo {volume} {07}},\
  \bibinfo {pages} {009} (\bibinfo {year} {2005})},\ \Eprint
  {http://arxiv.org/abs/hep-th/0502206} {arXiv:hep-th/0502206 [hep-th]}
  \BibitemShut {NoStop}%
\bibitem [{\citenamefont {Berti}\ \emph {et~al.}(2009)\citenamefont {Berti},
  \citenamefont {Cardoso},\ and\ \citenamefont {Starinets}}]{Berti:2009kk}%
  \BibitemOpen
  \bibfield  {author} {\bibinfo {author} {\bibfnamefont {E.}~\bibnamefont
  {Berti}}, \bibinfo {author} {\bibfnamefont {V.}~\bibnamefont {Cardoso}}, \
  and\ \bibinfo {author} {\bibfnamefont {A.~O.}\ \bibnamefont {Starinets}},\
  }\href {\doibase 10.1088/0264-9381/26/16/163001} {\bibfield  {journal}
  {\bibinfo  {journal} {Class. Quant. Grav.}\ }\textbf {\bibinfo {volume}
  {26}},\ \bibinfo {pages} {163001} (\bibinfo {year} {2009})},\ \Eprint
  {http://arxiv.org/abs/0905.2975} {arXiv:0905.2975 [gr-qc]} \BibitemShut
  {NoStop}%
\bibitem [{\citenamefont {{Zel'Dovich}}(1971)}]{1971JETPL..14..180Z}%
  \BibitemOpen
  \bibfield  {author} {\bibinfo {author} {\bibfnamefont {Y.~B.}\ \bibnamefont
  {{Zel'Dovich}}},\ }\href@noop {} {\bibfield  {journal} {\bibinfo  {journal}
  {Soviet Journal of Experimental and Theoretical Physics Letters}\ }\textbf
  {\bibinfo {volume} {14}},\ \bibinfo {pages} {180} (\bibinfo {year}
  {1971})}\BibitemShut {NoStop}%
\bibitem [{\citenamefont {Press}\ and\ \citenamefont
  {Teukolsky}(1972)}]{Press:1972zz}%
  \BibitemOpen
  \bibfield  {author} {\bibinfo {author} {\bibfnamefont {W.~H.}\ \bibnamefont
  {Press}}\ and\ \bibinfo {author} {\bibfnamefont {S.~A.}\ \bibnamefont
  {Teukolsky}},\ }\href {\doibase 10.1038/238211a0} {\bibfield  {journal}
  {\bibinfo  {journal} {Nature}\ }\textbf {\bibinfo {volume} {238}},\ \bibinfo
  {pages} {211} (\bibinfo {year} {1972})}\BibitemShut {NoStop}%
\bibitem [{\citenamefont {{Misner}}(1972)}]{1972BAPS...17..472M}%
  \BibitemOpen
  \bibfield  {author} {\bibinfo {author} {\bibfnamefont {C.}~\bibnamefont
  {{Misner}}},\ }\href@noop {} {\bibfield  {journal} {\bibinfo  {journal}
  {Bulletin of the American Physical Society}\ }\textbf {\bibinfo {volume}
  {17}},\ \bibinfo {pages} {472} (\bibinfo {year} {1972})}\BibitemShut
  {NoStop}%
\bibitem [{\citenamefont {{Starobinsky}}(1973)}]{staro1}%
  \BibitemOpen
  \bibfield  {author} {\bibinfo {author} {\bibfnamefont {A.~A.}\ \bibnamefont
  {{Starobinsky}}},\ }\href@noop {} {\bibfield  {journal} {\bibinfo  {journal}
  {Sov. Phys. JETP}\ }\textbf {\bibinfo {volume} {37}},\ \bibinfo {pages} {28}
  (\bibinfo {year} {1973})}\BibitemShut {NoStop}%
\bibitem [{\citenamefont {{Starobinsky}}\ and\ \citenamefont
  {{Churilov}}(1974)}]{staro2}%
  \BibitemOpen
  \bibfield  {author} {\bibinfo {author} {\bibfnamefont {A.~A.}\ \bibnamefont
  {{Starobinsky}}}\ and\ \bibinfo {author} {\bibfnamefont {S.~M.}\ \bibnamefont
  {{Churilov}}},\ }\href@noop {} {\bibfield  {journal} {\bibinfo  {journal}
  {Sov. Phys. JETP}\ }\textbf {\bibinfo {volume} {38}},\ \bibinfo {pages} {1}
  (\bibinfo {year} {1974})}\BibitemShut {NoStop}%
\bibitem [{\citenamefont {Torres}\ \emph {et~al.}(2017)\citenamefont {Torres},
  \citenamefont {Patrick}, \citenamefont {Coutant}, \citenamefont {Richartz},
  \citenamefont {Tedford},\ and\ \citenamefont {Weinfurtner}}]{Torres:2016iee}%
  \BibitemOpen
  \bibfield  {author} {\bibinfo {author} {\bibfnamefont {T.}~\bibnamefont
  {Torres}}, \bibinfo {author} {\bibfnamefont {S.}~\bibnamefont {Patrick}},
  \bibinfo {author} {\bibfnamefont {A.}~\bibnamefont {Coutant}}, \bibinfo
  {author} {\bibfnamefont {M.}~\bibnamefont {Richartz}}, \bibinfo {author}
  {\bibfnamefont {E.~W.}\ \bibnamefont {Tedford}}, \ and\ \bibinfo {author}
  {\bibfnamefont {S.}~\bibnamefont {Weinfurtner}},\ }\href {\doibase
  10.1038/nphys4151} {\bibfield  {journal} {\bibinfo  {journal} {Nature Phys.}\
  }\textbf {\bibinfo {volume} {13}},\ \bibinfo {pages} {833} (\bibinfo {year}
  {2017})},\ \Eprint {http://arxiv.org/abs/1612.06180} {arXiv:1612.06180
  [gr-qc]} \BibitemShut {NoStop}%
\bibitem [{\citenamefont {D'Antonio}\ \emph {et~al.}(2018)\citenamefont
  {D'Antonio} \emph {et~al.}}]{DAntonio:2018sff}%
  \BibitemOpen
  \bibfield  {author} {\bibinfo {author} {\bibfnamefont {S.}~\bibnamefont
  {D'Antonio}} \emph {et~al.},\ }\href@noop {} {\  (\bibinfo {year} {2018})},\
  \Eprint {http://arxiv.org/abs/1809.07202} {arXiv:1809.07202 [gr-qc]}
  \BibitemShut {NoStop}%
\bibitem [{\citenamefont {Teukolsky}(1973)}]{Teukolsky:1973ha}%
  \BibitemOpen
  \bibfield  {author} {\bibinfo {author} {\bibfnamefont {S.~A.}\ \bibnamefont
  {Teukolsky}},\ }\href {\doibase 10.1086/152444} {\bibfield  {journal}
  {\bibinfo  {journal} {Astrophys. J.}\ }\textbf {\bibinfo {volume} {185}},\
  \bibinfo {pages} {635} (\bibinfo {year} {1973})}\BibitemShut {NoStop}%
\bibitem [{\citenamefont {Sasaki}\ and\ \citenamefont
  {Tagoshi}(2003)}]{Sasaki:2003xr}%
  \BibitemOpen
  \bibfield  {author} {\bibinfo {author} {\bibfnamefont {M.}~\bibnamefont
  {Sasaki}}\ and\ \bibinfo {author} {\bibfnamefont {H.}~\bibnamefont
  {Tagoshi}},\ }\href {\doibase 10.12942/lrr-2003-6} {\bibfield  {journal}
  {\bibinfo  {journal} {Living Rev. Rel.}\ }\textbf {\bibinfo {volume} {6}},\
  \bibinfo {pages} {6} (\bibinfo {year} {2003})},\ \Eprint
  {http://arxiv.org/abs/gr-qc/0306120} {arXiv:gr-qc/0306120 [gr-qc]}
  \BibitemShut {NoStop}%
\bibitem [{\citenamefont {Yoshino}\ and\ \citenamefont
  {Kodama}(2014)}]{Yoshino:2013ofa}%
  \BibitemOpen
  \bibfield  {author} {\bibinfo {author} {\bibfnamefont {H.}~\bibnamefont
  {Yoshino}}\ and\ \bibinfo {author} {\bibfnamefont {H.}~\bibnamefont
  {Kodama}},\ }\href {\doibase 10.1093/ptep/ptu029} {\bibfield  {journal}
  {\bibinfo  {journal} {PTEP}\ }\textbf {\bibinfo {volume} {2014}},\ \bibinfo
  {pages} {043E02} (\bibinfo {year} {2014})},\ \Eprint
  {http://arxiv.org/abs/1312.2326} {arXiv:1312.2326 [gr-qc]} \BibitemShut
  {NoStop}%
\bibitem [{\citenamefont {Thorne}(1987)}]{Thorne:1987af}%
  \BibitemOpen
  \bibfield  {author} {\bibinfo {author} {\bibfnamefont {K.~S.}\ \bibnamefont
  {Thorne}},\ }\href@noop {} {\  (\bibinfo {year} {1987})}\BibitemShut
  {NoStop}%
\bibitem [{\citenamefont {Abbott}\ \emph
  {et~al.}(2016{\natexlab{a}})\citenamefont {Abbott} \emph
  {et~al.}}]{Abbott:2016blz}%
  \BibitemOpen
  \bibfield  {author} {\bibinfo {author} {\bibfnamefont {B.~P.}\ \bibnamefont
  {Abbott}} \emph {et~al.} (\bibinfo {collaboration} {Virgo, LIGO
  Scientific}),\ }\href {\doibase 10.1103/PhysRevLett.116.061102} {\bibfield
  {journal} {\bibinfo  {journal} {Phys. Rev. Lett.}\ }\textbf {\bibinfo
  {volume} {116}},\ \bibinfo {pages} {061102} (\bibinfo {year}
  {2016}{\natexlab{a}})},\ \Eprint {http://arxiv.org/abs/1602.03837}
  {arXiv:1602.03837 [gr-qc]} \BibitemShut {NoStop}%
\bibitem [{\citenamefont {Abbott}\ \emph {et~al.}(2017)\citenamefont {Abbott}
  \emph {et~al.}}]{Abbott:2017iws}%
  \BibitemOpen
  \bibfield  {author} {\bibinfo {author} {\bibfnamefont {B.~P.}\ \bibnamefont
  {Abbott}} \emph {et~al.} (\bibinfo {collaboration} {Virgo, LIGO
  Scientific}),\ }\href {\doibase 10.1103/PhysRevD.96.022001} {\bibfield
  {journal} {\bibinfo  {journal} {Phys. Rev.}\ }\textbf {\bibinfo {volume}
  {D96}},\ \bibinfo {pages} {022001} (\bibinfo {year} {2017})},\ \Eprint
  {http://arxiv.org/abs/1704.04628} {arXiv:1704.04628 [gr-qc]} \BibitemShut
  {NoStop}%
\bibitem [{\citenamefont {Calderon~Bustillo}\ \emph {et~al.}(2018)\citenamefont
  {Calderon~Bustillo}, \citenamefont {Salemi}, \citenamefont {Dal~Canton},\
  and\ \citenamefont {Jani}}]{CalderonBustillo:2017skv}%
  \BibitemOpen
  \bibfield  {author} {\bibinfo {author} {\bibfnamefont {J.}~\bibnamefont
  {Calderon~Bustillo}}, \bibinfo {author} {\bibfnamefont {F.}~\bibnamefont
  {Salemi}}, \bibinfo {author} {\bibfnamefont {T.}~\bibnamefont {Dal~Canton}},
  \ and\ \bibinfo {author} {\bibfnamefont {K.~P.}\ \bibnamefont {Jani}},\
  }\href {\doibase 10.1103/PhysRevD.97.024016} {\bibfield  {journal} {\bibinfo
  {journal} {Phys. Rev.}\ }\textbf {\bibinfo {volume} {D97}},\ \bibinfo {pages}
  {024016} (\bibinfo {year} {2018})},\ \Eprint
  {http://arxiv.org/abs/1711.02009} {arXiv:1711.02009 [gr-qc]} \BibitemShut
  {NoStop}%
\bibitem [{\citenamefont {Regimbau}(2011)}]{Regimbau:2011rp}%
  \BibitemOpen
  \bibfield  {author} {\bibinfo {author} {\bibfnamefont {T.}~\bibnamefont
  {Regimbau}},\ }\href {\doibase 10.1088/1674-4527/11/4/001} {\bibfield
  {journal} {\bibinfo  {journal} {Res. Astron. Astrophys.}\ }\textbf {\bibinfo
  {volume} {11}},\ \bibinfo {pages} {369} (\bibinfo {year} {2011})},\ \Eprint
  {http://arxiv.org/abs/1101.2762} {arXiv:1101.2762 [astro-ph.CO]} \BibitemShut
  {NoStop}%
\bibitem [{\citenamefont {Thrane}\ and\ \citenamefont
  {Romano}(2013)}]{Thrane:2013oya}%
  \BibitemOpen
  \bibfield  {author} {\bibinfo {author} {\bibfnamefont {E.}~\bibnamefont
  {Thrane}}\ and\ \bibinfo {author} {\bibfnamefont {J.~D.}\ \bibnamefont
  {Romano}},\ }\href {\doibase 10.1103/PhysRevD.88.124032} {\bibfield
  {journal} {\bibinfo  {journal} {Phys. Rev.}\ }\textbf {\bibinfo {volume}
  {D88}},\ \bibinfo {pages} {124032} (\bibinfo {year} {2013})},\ \Eprint
  {http://arxiv.org/abs/1310.5300} {arXiv:1310.5300 [astro-ph.IM]} \BibitemShut
  {NoStop}%
\bibitem [{\citenamefont {Cornish}\ and\ \citenamefont
  {Robson}(2017)}]{Cornish:2017vip}%
  \BibitemOpen
  \bibfield  {author} {\bibinfo {author} {\bibfnamefont {N.}~\bibnamefont
  {Cornish}}\ and\ \bibinfo {author} {\bibfnamefont {T.}~\bibnamefont
  {Robson}},\ }\bibfield  {booktitle} {\emph {\bibinfo {booktitle}
  {{Proceedings, 11th International LISA Symposium: Zurich, Switzerland,
  September 5-9, 2016}}},\ }\href {\doibase 10.1088/1742-6596/840/1/012024}
  {\bibfield  {journal} {\bibinfo  {journal} {J. Phys. Conf. Ser.}\ }\textbf
  {\bibinfo {volume} {840}},\ \bibinfo {pages} {012024} (\bibinfo {year}
  {2017})},\ \Eprint {http://arxiv.org/abs/1703.09858} {arXiv:1703.09858
  [astro-ph.IM]} \BibitemShut {NoStop}%
\bibitem [{\citenamefont {Hartwig}\ \emph {et~al.}(2016)\citenamefont
  {Hartwig}, \citenamefont {Volonteri}, \citenamefont {Bromm}, \citenamefont
  {Klessen}, \citenamefont {Barausse}, \citenamefont {Magg},\ and\
  \citenamefont {Stacy}}]{Hartwig:2016nde}%
  \BibitemOpen
  \bibfield  {author} {\bibinfo {author} {\bibfnamefont {T.}~\bibnamefont
  {Hartwig}}, \bibinfo {author} {\bibfnamefont {M.}~\bibnamefont {Volonteri}},
  \bibinfo {author} {\bibfnamefont {V.}~\bibnamefont {Bromm}}, \bibinfo
  {author} {\bibfnamefont {R.~S.}\ \bibnamefont {Klessen}}, \bibinfo {author}
  {\bibfnamefont {E.}~\bibnamefont {Barausse}}, \bibinfo {author}
  {\bibfnamefont {M.}~\bibnamefont {Magg}}, \ and\ \bibinfo {author}
  {\bibfnamefont {A.}~\bibnamefont {Stacy}},\ }\href {\doibase
  10.1093/mnrasl/slw074} {\bibfield  {journal} {\bibinfo  {journal} {Mon. Not.
  Roy. Astron. Soc.}\ }\textbf {\bibinfo {volume} {460}},\ \bibinfo {pages}
  {L74} (\bibinfo {year} {2016})},\ \Eprint {http://arxiv.org/abs/1603.05655}
  {arXiv:1603.05655 [astro-ph.GA]} \BibitemShut {NoStop}%
\bibitem [{\citenamefont {Aghanim}\ \emph {et~al.}(2018)\citenamefont {Aghanim}
  \emph {et~al.}}]{Aghanim:2018eyx}%
  \BibitemOpen
  \bibfield  {author} {\bibinfo {author} {\bibfnamefont {N.}~\bibnamefont
  {Aghanim}} \emph {et~al.} (\bibinfo {collaboration} {Planck}),\ }\href@noop
  {} {\  (\bibinfo {year} {2018})},\ \Eprint {http://arxiv.org/abs/1807.06209}
  {arXiv:1807.06209 [astro-ph.CO]} \BibitemShut {NoStop}%
\bibitem [{\citenamefont {Dvorkin}\ \emph {et~al.}(2016)\citenamefont
  {Dvorkin}, \citenamefont {Vangioni}, \citenamefont {Silk}, \citenamefont
  {Uzan},\ and\ \citenamefont {Olive}}]{Dvorkin:2016wac}%
  \BibitemOpen
  \bibfield  {author} {\bibinfo {author} {\bibfnamefont {I.}~\bibnamefont
  {Dvorkin}}, \bibinfo {author} {\bibfnamefont {E.}~\bibnamefont {Vangioni}},
  \bibinfo {author} {\bibfnamefont {J.}~\bibnamefont {Silk}}, \bibinfo {author}
  {\bibfnamefont {J.-P.}\ \bibnamefont {Uzan}}, \ and\ \bibinfo {author}
  {\bibfnamefont {K.~A.}\ \bibnamefont {Olive}},\ }\href {\doibase
  10.1093/mnras/stw1477} {\bibfield  {journal} {\bibinfo  {journal} {Mon. Not.
  Roy. Astron. Soc.}\ }\textbf {\bibinfo {volume} {461}},\ \bibinfo {pages}
  {3877} (\bibinfo {year} {2016})},\ \Eprint {http://arxiv.org/abs/1604.04288}
  {arXiv:1604.04288 [astro-ph.HE]} \BibitemShut {NoStop}%
\bibitem [{\citenamefont {{The LIGO Scientific Collaboration and The Virgo
  Collaboration}}(2018{\natexlab{b}})}]{LIGOScientific:2018jsj}%
  \BibitemOpen
  \bibfield  {author} {\bibinfo {author} {\bibnamefont {{The LIGO Scientific
  Collaboration and The Virgo Collaboration}}},\ }\href@noop {} {\  (\bibinfo
  {year} {2018}{\natexlab{b}})},\ \Eprint {http://arxiv.org/abs/1811.12940}
  {arXiv:1811.12940 [astro-ph.HE]} \BibitemShut {NoStop}%
\bibitem [{\citenamefont {Abbott}\ \emph
  {et~al.}(2016{\natexlab{b}})\citenamefont {Abbott} \emph
  {et~al.}}]{TheLIGOScientific:2016pea}%
  \BibitemOpen
  \bibfield  {author} {\bibinfo {author} {\bibfnamefont {B.~P.}\ \bibnamefont
  {Abbott}} \emph {et~al.} (\bibinfo {collaboration} {Virgo, LIGO
  Scientific}),\ }\href {\doibase 10.1103/PhysRevX.6.041015,
  10.1103/PhysRevX.8.039903} {\bibfield  {journal} {\bibinfo  {journal} {Phys.
  Rev.}\ }\textbf {\bibinfo {volume} {X6}},\ \bibinfo {pages} {041015}
  (\bibinfo {year} {2016}{\natexlab{b}})},\ \bibinfo {note} {[Erratum: Phys.
  Rev.X8,no.3,039903(2018)]},\ \Eprint {http://arxiv.org/abs/1606.04856}
  {arXiv:1606.04856 [gr-qc]} \BibitemShut {NoStop}%
\bibitem [{\citenamefont {Kovetz}\ \emph {et~al.}(2017)\citenamefont {Kovetz},
  \citenamefont {Cholis}, \citenamefont {Breysse},\ and\ \citenamefont
  {Kamionkowski}}]{Kovetz:2016kpi}%
  \BibitemOpen
  \bibfield  {author} {\bibinfo {author} {\bibfnamefont {E.~D.}\ \bibnamefont
  {Kovetz}}, \bibinfo {author} {\bibfnamefont {I.}~\bibnamefont {Cholis}},
  \bibinfo {author} {\bibfnamefont {P.~C.}\ \bibnamefont {Breysse}}, \ and\
  \bibinfo {author} {\bibfnamefont {M.}~\bibnamefont {Kamionkowski}},\ }\href
  {\doibase 10.1103/PhysRevD.95.103010} {\bibfield  {journal} {\bibinfo
  {journal} {Phys. Rev.}\ }\textbf {\bibinfo {volume} {D95}},\ \bibinfo {pages}
  {103010} (\bibinfo {year} {2017})},\ \Eprint
  {http://arxiv.org/abs/1611.01157} {arXiv:1611.01157 [astro-ph.CO]}
  \BibitemShut {NoStop}%
\bibitem [{\citenamefont {Belczynski}\ \emph {et~al.}(2016)\citenamefont
  {Belczynski} \emph {et~al.}}]{Belczynski:2016jno}%
  \BibitemOpen
  \bibfield  {author} {\bibinfo {author} {\bibfnamefont {K.}~\bibnamefont
  {Belczynski}} \emph {et~al.},\ }\href {\doibase 10.1051/0004-6361/201628980}
  {\bibfield  {journal} {\bibinfo  {journal} {Astron. Astrophys.}\ }\textbf
  {\bibinfo {volume} {594}},\ \bibinfo {pages} {A97} (\bibinfo {year}
  {2016})},\ \Eprint {http://arxiv.org/abs/1607.03116} {arXiv:1607.03116
  [astro-ph.HE]} \BibitemShut {NoStop}%
\bibitem [{\citenamefont {Gerosa}\ and\ \citenamefont
  {Berti}(2017)}]{Gerosa:2017kvu}%
  \BibitemOpen
  \bibfield  {author} {\bibinfo {author} {\bibfnamefont {D.}~\bibnamefont
  {Gerosa}}\ and\ \bibinfo {author} {\bibfnamefont {E.}~\bibnamefont {Berti}},\
  }\href {\doibase 10.1103/PhysRevD.95.124046} {\bibfield  {journal} {\bibinfo
  {journal} {Phys. Rev.}\ }\textbf {\bibinfo {volume} {D95}},\ \bibinfo {pages}
  {124046} (\bibinfo {year} {2017})},\ \Eprint
  {http://arxiv.org/abs/1703.06223} {arXiv:1703.06223 [gr-qc]} \BibitemShut
  {NoStop}%
\bibitem [{\citenamefont {Fishbach}\ \emph {et~al.}(2017)\citenamefont
  {Fishbach}, \citenamefont {Holz},\ and\ \citenamefont
  {Farr}}]{Fishbach:2017dwv}%
  \BibitemOpen
  \bibfield  {author} {\bibinfo {author} {\bibfnamefont {M.}~\bibnamefont
  {Fishbach}}, \bibinfo {author} {\bibfnamefont {D.~E.}\ \bibnamefont {Holz}},
  \ and\ \bibinfo {author} {\bibfnamefont {B.}~\bibnamefont {Farr}},\ }\href
  {\doibase 10.3847/2041-8213/aa7045} {\bibfield  {journal} {\bibinfo
  {journal} {Astrophys. J.}\ }\textbf {\bibinfo {volume} {840}},\ \bibinfo
  {pages} {L24} (\bibinfo {year} {2017})},\ \Eprint
  {http://arxiv.org/abs/1703.06869} {arXiv:1703.06869 [astro-ph.HE]}
  \BibitemShut {NoStop}%
\bibitem [{\citenamefont {Rodriguez}\ \emph
  {et~al.}(2016{\natexlab{a}})\citenamefont {Rodriguez}, \citenamefont
  {Chatterjee},\ and\ \citenamefont {Rasio}}]{Rodriguez:2016kxx}%
  \BibitemOpen
  \bibfield  {author} {\bibinfo {author} {\bibfnamefont {C.~L.}\ \bibnamefont
  {Rodriguez}}, \bibinfo {author} {\bibfnamefont {S.}~\bibnamefont
  {Chatterjee}}, \ and\ \bibinfo {author} {\bibfnamefont {F.~A.}\ \bibnamefont
  {Rasio}},\ }\href {\doibase 10.1103/PhysRevD.93.084029} {\bibfield  {journal}
  {\bibinfo  {journal} {Phys. Rev.}\ }\textbf {\bibinfo {volume} {D93}},\
  \bibinfo {pages} {084029} (\bibinfo {year} {2016}{\natexlab{a}})},\ \Eprint
  {http://arxiv.org/abs/1602.02444} {arXiv:1602.02444 [astro-ph.HE]}
  \BibitemShut {NoStop}%
\bibitem [{\citenamefont {Samsing}\ and\ \citenamefont
  {D'Orazio}(2018)}]{Samsing:2018isx}%
  \BibitemOpen
  \bibfield  {author} {\bibinfo {author} {\bibfnamefont {J.}~\bibnamefont
  {Samsing}}\ and\ \bibinfo {author} {\bibfnamefont {D.~J.}\ \bibnamefont
  {D'Orazio}},\ }\href {\doibase 10.1093/mnras/sty2334} {\  (\bibinfo {year}
  {2018}),\ 10.1093/mnras/sty2334},\ \Eprint {http://arxiv.org/abs/1804.06519}
  {arXiv:1804.06519 [astro-ph.HE]} \BibitemShut {NoStop}%
\bibitem [{\citenamefont {D'Orazio}\ and\ \citenamefont
  {Samsing}(2018)}]{DOrazio:2018jnv}%
  \BibitemOpen
  \bibfield  {author} {\bibinfo {author} {\bibfnamefont {D.~J.}\ \bibnamefont
  {D'Orazio}}\ and\ \bibinfo {author} {\bibfnamefont {J.}~\bibnamefont
  {Samsing}},\ }\href {\doibase 10.1093/mnras/sty2568} {\  (\bibinfo {year}
  {2018}),\ 10.1093/mnras/sty2568},\ \Eprint {http://arxiv.org/abs/1805.06194}
  {arXiv:1805.06194 [astro-ph.HE]} \BibitemShut {NoStop}%
\bibitem [{\citenamefont {Belczynski}\ \emph {et~al.}(2017)\citenamefont
  {Belczynski}, \citenamefont {Ryu}, \citenamefont {Perna}, \citenamefont
  {Berti}, \citenamefont {Tanaka},\ and\ \citenamefont
  {Bulik}}]{Belczynski:2016ieo}%
  \BibitemOpen
  \bibfield  {author} {\bibinfo {author} {\bibfnamefont {K.}~\bibnamefont
  {Belczynski}}, \bibinfo {author} {\bibfnamefont {T.}~\bibnamefont {Ryu}},
  \bibinfo {author} {\bibfnamefont {R.}~\bibnamefont {Perna}}, \bibinfo
  {author} {\bibfnamefont {E.}~\bibnamefont {Berti}}, \bibinfo {author}
  {\bibfnamefont {T.~L.}\ \bibnamefont {Tanaka}}, \ and\ \bibinfo {author}
  {\bibfnamefont {T.}~\bibnamefont {Bulik}},\ }\href {\doibase
  10.1093/mnras/stx1759} {\bibfield  {journal} {\bibinfo  {journal} {Mon. Not.
  Roy. Astron. Soc.}\ }\textbf {\bibinfo {volume} {471}},\ \bibinfo {pages}
  {4702} (\bibinfo {year} {2017})},\ \Eprint {http://arxiv.org/abs/1612.01524}
  {arXiv:1612.01524 [astro-ph.HE]} \BibitemShut {NoStop}%
\bibitem [{\citenamefont {Rodriguez}\ \emph
  {et~al.}(2016{\natexlab{b}})\citenamefont {Rodriguez}, \citenamefont {Zevin},
  \citenamefont {Pankow}, \citenamefont {Kalogera},\ and\ \citenamefont
  {Rasio}}]{Rodriguez:2016vmx}%
  \BibitemOpen
  \bibfield  {author} {\bibinfo {author} {\bibfnamefont {C.~L.}\ \bibnamefont
  {Rodriguez}}, \bibinfo {author} {\bibfnamefont {M.}~\bibnamefont {Zevin}},
  \bibinfo {author} {\bibfnamefont {C.}~\bibnamefont {Pankow}}, \bibinfo
  {author} {\bibfnamefont {V.}~\bibnamefont {Kalogera}}, \ and\ \bibinfo
  {author} {\bibfnamefont {F.~A.}\ \bibnamefont {Rasio}},\ }\href {\doibase
  10.3847/2041-8205/832/1/L2} {\bibfield  {journal} {\bibinfo  {journal}
  {Astrophys. J.}\ }\textbf {\bibinfo {volume} {832}},\ \bibinfo {pages} {L2}
  (\bibinfo {year} {2016}{\natexlab{b}})},\ \Eprint
  {http://arxiv.org/abs/1609.05916} {arXiv:1609.05916 [astro-ph.HE]}
  \BibitemShut {NoStop}%
\bibitem [{\citenamefont {Berti}\ and\ \citenamefont
  {Volonteri}(2008)}]{Berti:2008af}%
  \BibitemOpen
  \bibfield  {author} {\bibinfo {author} {\bibfnamefont {E.}~\bibnamefont
  {Berti}}\ and\ \bibinfo {author} {\bibfnamefont {M.}~\bibnamefont
  {Volonteri}},\ }\href {\doibase 10.1086/590379} {\bibfield  {journal}
  {\bibinfo  {journal} {Astrophys. J.}\ }\textbf {\bibinfo {volume} {684}},\
  \bibinfo {pages} {822} (\bibinfo {year} {2008})},\ \Eprint
  {http://arxiv.org/abs/0802.0025} {arXiv:0802.0025 [astro-ph]} \BibitemShut
  {NoStop}%
\bibitem [{\citenamefont {Gerosa}\ \emph {et~al.}(2018)\citenamefont {Gerosa},
  \citenamefont {Berti}, \citenamefont {O'Shaughnessy}, \citenamefont
  {Belczynski}, \citenamefont {Kesden}, \citenamefont {Wysocki},\ and\
  \citenamefont {Gladysz}}]{Gerosa:2018wbw}%
  \BibitemOpen
  \bibfield  {author} {\bibinfo {author} {\bibfnamefont {D.}~\bibnamefont
  {Gerosa}}, \bibinfo {author} {\bibfnamefont {E.}~\bibnamefont {Berti}},
  \bibinfo {author} {\bibfnamefont {R.}~\bibnamefont {O'Shaughnessy}}, \bibinfo
  {author} {\bibfnamefont {K.}~\bibnamefont {Belczynski}}, \bibinfo {author}
  {\bibfnamefont {M.}~\bibnamefont {Kesden}}, \bibinfo {author} {\bibfnamefont
  {D.}~\bibnamefont {Wysocki}}, \ and\ \bibinfo {author} {\bibfnamefont
  {W.}~\bibnamefont {Gladysz}},\ }\href {\doibase 10.1103/PhysRevD.98.084036}
  {\bibfield  {journal} {\bibinfo  {journal} {Phys. Rev.}\ }\textbf {\bibinfo
  {volume} {D98}},\ \bibinfo {pages} {084036} (\bibinfo {year} {2018})},\
  \Eprint {http://arxiv.org/abs/1808.02491} {arXiv:1808.02491 [astro-ph.HE]}
  \BibitemShut {NoStop}%
\bibitem [{\citenamefont {Barausse}\ and\ \citenamefont
  {Rezzolla}(2009)}]{Barausse:2009uz}%
  \BibitemOpen
  \bibfield  {author} {\bibinfo {author} {\bibfnamefont {E.}~\bibnamefont
  {Barausse}}\ and\ \bibinfo {author} {\bibfnamefont {L.}~\bibnamefont
  {Rezzolla}},\ }\href {\doibase 10.1088/0004-637X/704/1/L40} {\bibfield
  {journal} {\bibinfo  {journal} {Astrophys. J.}\ }\textbf {\bibinfo {volume}
  {704}},\ \bibinfo {pages} {L40} (\bibinfo {year} {2009})},\ \Eprint
  {http://arxiv.org/abs/0904.2577} {arXiv:0904.2577 [gr-qc]} \BibitemShut
  {NoStop}%
\bibitem [{\citenamefont {Barausse}\ \emph {et~al.}(2012)\citenamefont
  {Barausse}, \citenamefont {Morozova},\ and\ \citenamefont
  {Rezzolla}}]{Barausse:2012qz}%
  \BibitemOpen
  \bibfield  {author} {\bibinfo {author} {\bibfnamefont {E.}~\bibnamefont
  {Barausse}}, \bibinfo {author} {\bibfnamefont {V.}~\bibnamefont {Morozova}},
  \ and\ \bibinfo {author} {\bibfnamefont {L.}~\bibnamefont {Rezzolla}},\
  }\href {\doibase 10.1088/0004-637X/786/1/76, 10.1088/0004-637X/758/1/63}
  {\bibfield  {journal} {\bibinfo  {journal} {Astrophys. J.}\ }\textbf
  {\bibinfo {volume} {758}},\ \bibinfo {pages} {63} (\bibinfo {year} {2012})},\
  \bibinfo {note} {[Erratum: Astrophys. J.786,76(2014)]},\ \Eprint
  {http://arxiv.org/abs/1206.3803} {arXiv:1206.3803 [gr-qc]} \BibitemShut
  {NoStop}%
\bibitem [{\citenamefont {Buonanno}\ \emph {et~al.}(2008)\citenamefont
  {Buonanno}, \citenamefont {Kidder},\ and\ \citenamefont
  {Lehner}}]{Buonanno:2007sv}%
  \BibitemOpen
  \bibfield  {author} {\bibinfo {author} {\bibfnamefont {A.}~\bibnamefont
  {Buonanno}}, \bibinfo {author} {\bibfnamefont {L.~E.}\ \bibnamefont
  {Kidder}}, \ and\ \bibinfo {author} {\bibfnamefont {L.}~\bibnamefont
  {Lehner}},\ }\href {\doibase 10.1103/PhysRevD.77.026004} {\bibfield
  {journal} {\bibinfo  {journal} {Phys. Rev.}\ }\textbf {\bibinfo {volume}
  {D77}},\ \bibinfo {pages} {026004} (\bibinfo {year} {2008})},\ \Eprint
  {http://arxiv.org/abs/0709.3839} {arXiv:0709.3839 [astro-ph]} \BibitemShut
  {NoStop}%
\bibitem [{\citenamefont {Berti}\ \emph {et~al.}(2007)\citenamefont {Berti},
  \citenamefont {Cardoso}, \citenamefont {Gonzalez}, \citenamefont {Sperhake},
  \citenamefont {Hannam}, \citenamefont {Husa},\ and\ \citenamefont
  {Brugmann}}]{Berti:2007fi}%
  \BibitemOpen
  \bibfield  {author} {\bibinfo {author} {\bibfnamefont {E.}~\bibnamefont
  {Berti}}, \bibinfo {author} {\bibfnamefont {V.}~\bibnamefont {Cardoso}},
  \bibinfo {author} {\bibfnamefont {J.~A.}\ \bibnamefont {Gonzalez}}, \bibinfo
  {author} {\bibfnamefont {U.}~\bibnamefont {Sperhake}}, \bibinfo {author}
  {\bibfnamefont {M.}~\bibnamefont {Hannam}}, \bibinfo {author} {\bibfnamefont
  {S.}~\bibnamefont {Husa}}, \ and\ \bibinfo {author} {\bibfnamefont
  {B.}~\bibnamefont {Brugmann}},\ }\href {\doibase 10.1103/PhysRevD.76.064034}
  {\bibfield  {journal} {\bibinfo  {journal} {Phys. Rev.}\ }\textbf {\bibinfo
  {volume} {D76}},\ \bibinfo {pages} {064034} (\bibinfo {year} {2007})},\
  \Eprint {http://arxiv.org/abs/gr-qc/0703053} {arXiv:gr-qc/0703053 [GR-QC]}
  \BibitemShut {NoStop}%
\bibitem [{\citenamefont {East}(2017)}]{East:2017mrj}%
  \BibitemOpen
  \bibfield  {author} {\bibinfo {author} {\bibfnamefont {W.~E.}\ \bibnamefont
  {East}},\ }\href {\doibase 10.1103/PhysRevD.96.024004} {\bibfield  {journal}
  {\bibinfo  {journal} {Phys. Rev.}\ }\textbf {\bibinfo {volume} {D96}},\
  \bibinfo {pages} {024004} (\bibinfo {year} {2017})},\ \Eprint
  {http://arxiv.org/abs/1705.01544} {arXiv:1705.01544 [gr-qc]} \BibitemShut
  {NoStop}%
\bibitem [{\citenamefont {Brito}\ \emph {et~al.}(2013)\citenamefont {Brito},
  \citenamefont {Cardoso},\ and\ \citenamefont {Pani}}]{Brito:2013wya}%
  \BibitemOpen
  \bibfield  {author} {\bibinfo {author} {\bibfnamefont {R.}~\bibnamefont
  {Brito}}, \bibinfo {author} {\bibfnamefont {V.}~\bibnamefont {Cardoso}}, \
  and\ \bibinfo {author} {\bibfnamefont {P.}~\bibnamefont {Pani}},\ }\href
  {\doibase 10.1103/PhysRevD.88.023514} {\bibfield  {journal} {\bibinfo
  {journal} {Phys. Rev.}\ }\textbf {\bibinfo {volume} {D88}},\ \bibinfo {pages}
  {023514} (\bibinfo {year} {2013})},\ \Eprint {http://arxiv.org/abs/1304.6725}
  {arXiv:1304.6725 [gr-qc]} \BibitemShut {NoStop}%
\end{thebibliography}%

\end{document}